\documentclass[twocolumn]{aastex63}
\usepackage{amsmath}
\usepackage{multirow}
\usepackage{amssymb}
\usepackage{graphicx}
\usepackage{color}
\usepackage{float}
\usepackage{comment}

\definecolor{darkgreen}{rgb}{0,0.35,0}
\usepackage{CJK}

\newcommand{\unit}[1]{\,{\rm #1}}
\newcommand{\cs}{c_{\rm s}}
\newcommand{\kB}{k_{\textsc{b}}}
\newcommand{\prg}{\Pi}

\shorttitle{RHD Simulations of Gravitational Instability}
\shortauthors{Chen et al.}

\begin{document}
\begin{CJK*}{UTF8}{gbsn}

\title{3D Radiation Hydrodynamic Simulations of Gravitational Instability in AGN Accretion Disks: Effects of Radiation Pressure}

\author[0000-0003-3792-2888]{Yi-Xian Chen (陈逸贤)}
\email{yc9993@princeton.edu}
\affiliation{Department of Astrophysical Sciences, Princeton University, Princeton, NJ 08544, USA}
\author[0000-0002-2624-3399]{Yan-Fei Jiang (姜燕飞)}
\affiliation{Center for Computational Astrophysics, Flatiron Institute, New York, NY 10010, USA}
\author[0000-0002-6710-7748]{Jeremy Goodman}
\affiliation{Department of Astrophysical Sciences, Princeton University, Princeton, NJ 08544, USA}
\author[0000-0002-0509-9113]{Eve C. Ostriker}
\affiliation{Department of Astrophysical Sciences, Princeton University, Princeton, NJ 08544, USA}
\begin{abstract}
We perform 3D radiation hydrodynamic local shearing box simulations to 
study the outcome of gravitational instability (GI) 
in optically thick Active Galactic Nuclei (AGN) 
accretion disks. 
GI develops when the Toomre parameter $Q_T\lesssim 1$, 
and may lead to turbulent heating that balances radiative cooling. 
However, when radiative cooling is too efficient, the disk may undergo runaway gravitational fragmentation. 
In the fully gas-pressure-dominated case, we confirm the classical result that such a thermal balance holds when the Shakura-Sunyaev viscosity parameter ($\alpha$) due to the gravitationally-driven turbulence is $\lesssim 0.2$, corresponding to dimensionless cooling times $\Omega t_{\rm cool} \gtrsim 5$.  As the fraction of support by radiation pressure increases, the disk becomes more prone to fragmentation, with a reduced (increased) critical value of   $\alpha$ ($\Omega t_{\rm cool}$).  
The effect is already significant when 
the radiation pressure exceeds 10\% of the gas pressure, 
while fully radiation-pressure-dominated disks 
fragment at $t_{\rm cool}\lesssim 50\Omega^{-1}$.
The latter translates to 
a maximum 
turbulence level $\alpha \lesssim 0.02$, 
comparable to that generated by Magnetorotational Instability (MRI).
Our results suggest that gravitationally unstable ($Q_T\sim 1$) outer regions of AGN disks with significant radiation pressure 
(likely for high/near-Eddington accretion rates) should always fragment into stars, and perhaps black holes.
\end{abstract}

\section{Introduction}

Supermassive black holes (SMBHs) have been found in the centers of most massive galaxies \citep[see][for a review]{KormendyHo2013}. 
They typically have masses ranging from $10^6 M_{\odot} - 10^9 M_{\odot}$, 
and harbor accretion disks which provide power to active galactic nuclei (AGN) and quasars \citep{Lynden-Bell1969}. 


The outer regions of such disks are thought to be heated at least partially by gravito-turbulence, 
a process extensively studied in contexts of protoplanetary disks (PPDs) as well as AGN accretion disks \citep[e.g.][]{Gammie2001,Johnson2003,Rice2003,Rice2005}.
In a sufficiently extended standard thin accretion disk model \citep{Shakura1973} with dimensionless viscosity parameter $\alpha<1$ 
supporting a radially constant mass accretion rate ($\dot M$)---perhaps due to Magnetorotational Instability (MRI) \citep{Balbus1991}---there exists a self-gravitating radius $r_{\rm sg}$ 
beyond which the Toomre parameter $Q_T$ \citep{1964ApJ...139.1217T} drops below unity. 
Beyond this radius, 
which is typically around 0.01-0.1 pc for Eddington accretion rates and $M_{\rm SMBH}\sim 10^{8}M_\odot$, the disk becomes gravitationally unstable.  Disks may however be able to self-regulate at $Q_T \gtrsim 1$, provided the mass-feeding rate is low enough that accretion and heating can be sustained with a gravitationally-produced  $\alpha_{\rm GI}$ not so large as to induce fragmentation. 
Analytical models have been applied to describe this region 
as a constant-$Q_T$ disk in which the steady state turbulence from gravitational instability $\alpha_{\rm GI}$ is an explicit function of distance to the SMBH, $r$, 
parameterized by the accretion rate $\dot{M}$ and the value of $Q_T$  \citep{Goodman2003,GoodmanTan2004,SirkoGoodman2003,Levin2007}. 


In the case where both external heating and other sources of turbulence are weak, turbulence generated by gravitational instability must extract energy from the mean shear at a rate sufficient to offset radiative cooling, which requires $\alpha_{\rm GI} \sim (\Omega t_{\rm cool})^{-1}\equiv (\tau_{\rm cool})^{-1}$ for local cooling time $t_{\rm cool}$ and orbital frequency $\Omega$.
Extensive simulations in the gas pressure dominated regime \citep[e.g.][]{Gammie2001,Johnson2003,Rice2003} have led to the conclusion that if $\tau_{\rm cool} \lesssim 3-5 $, a statistical steady state cannot be achieved, and the disk fragments.
The exact value of the critical cooling time depends upon the equation of state \citep{Rice2005}, 
and still remains somewhat uncertain theoretically since  
numerical parameters such as resolution and integration time also appear to affect outcomes \citep{Paardekooper2012}.
The evolution and ultimate masses of the fragments is still under debate \citep{GoodmanTan2004,Levin2007}.

As we move further out in radius in a constant-$Q_T$ disk, radiative cooling becomes more efficient, 
and $\tau_{\rm cool}$ continues to decrease until $\tau_{\rm cool}\ll 1$ 
(especially when temperature drops to around 2000K and opacity becomes extremely small). 
Beyond this point disk fragmentation would seem to be inevitable, 
according to the simple $\tau_{\rm cool} \lesssim 3$ criterion derived from gas-pressure-dominated simulations.
Of course, in this situation, 
fragmentation is likely to lead to intense star formation, 
and the heat input from these stars may raise the sound speed and hence $Q_T$; 
also, 
enhanced  angular momentum transport via magnetized winds may further help to stabilize the disk by lowering the surface density required for a given accretion rate \citep{Goodman2003,TQM}. 


In previous work, 
the influence of radiation pressure on the critical cooling timescale has rarely been studied, 
and uncertainty in the fragmentation condition prevents us from gaining a deeper understanding of how a gravito-turbulent, 
sub-parsec and optically thick region bridges the gap between the luminous inner accreting regions and the outer star-forming regions of an AGN disk. 
For moderate $\dot{M} \gtrsim 1 M_\odot /$yr, 
radiation pressure dominates the total pressure within the gravitationally unstable disk region \citep{Goodman2003,GoodmanTan2004}. 
A radiation-supported disk has a local effective adiabatic index of 4/3 and is much more prone to GI fragmentation 
than disk supported by completely ionized hydrogen pressure 
(analogous to some soft spots in the equation of state for partially dissociated hydrogen, 
which induce GI fragmentation in irradiated PPDs \citep{HiroseShi2017,HiroseShi2019}).
Indeed, 
simulations from \citet{Jiang11} have shown that a high radiation-pressure fraction is able to push the critical $\tau_{\rm cool}$ to values much larger than the order-unity critical value that holds in the gas-pressure-dominated regime. 
In other words, it is more difficult for a radiation-dominated disk to support quasi-steady turbulence whose dissipation balances cooling  without runaway fragmentation.


The previous work of  \citet{Jiang11} employed shearing-sheet 2D simulations with a polytropic prescription for vertical hydrodynamical structure, and assumed an artificial prescription for radiative cooling $\propto T^4$. 
While the results from these simulations   were informative, this was primarily by way of identifying qualitative trends, given the highly simplified numerical treatment.  In particular, since the cooling rate and the disk's thermal response are key quantities that control GI, realistic thermodynamics and proper treatment of three-dimensional structure and radiation-gas interactions are essential to studying the non-linear evolution of GI in realistic AGN disks.

In this paper, 
we perform local shearing box simulations in 3D coupled with full radiation transport calculation, using the 
state-of-the-art implicit radiation module of \texttt{Athena++} \citep{Jiang2021}. 
Due to the high computational expense of radiative  transfer calculations and the long thermal timescales in high optical depth environments, 
we focus on tracing the fragmentation boundary and only follow fragmenting cases up to the point of runaway collapse. 
We do not attempt to estimate the final masses of fragments, as this may involve prolonged processes of inflow and outflow, coalescence, and perhaps nuclear burning, and would probably require subgrid models to avoid extreme demands on numerical resolution.
We follow quasi-steady cases for a few hundred dynamical timescales. 

This paper is organized as follows: In \S \ref{sec:formulation}, 
we lay out the theoretical framework for some basic scalings of thermodynamic quantities in the local parameter space assuming quasi-steady state, 
as well as their connection to a $Q_T\sim 1$ global disk model. In
\S \ref{sec:setup} we introduce our numerical setup for radiation hydrodynamic (RHD) simulations, 
including initial conditions, 
boundary conditions, 
diagnostics, and 
methods of determining fragmentation. 
Our results are presented and summarized in \S \ref{sec:results}.  Our models include both cases that result in  quasi-steady turbulence and those that undergo fragmentation, and both gas pressure and radiation pressure dominated regimes. 
We discuss the implication of our results and future prospects in \S \ref{sec:discussion}.



\section{Theoretical Formulation and Basic Scalings}
\label{sec:formulation}

As an idealized local model of an accretion disk, 
one of the main advantages of the shearing box is that it is defined by a small number of control parameters that should remain constant as the system evolves.
For simulations of gravito-turbulence, 
the principle control parameters of a Keplerian (i.e. shearing factor $q \equiv -d\ln\Omega/d\ln r=3/2$) shearing box are the surface density of mass, $\Sigma$, 
and the rotation rate, 
$\Omega$.
In combination with Newton's constant $G$, 
we can construct dynamically relevant units of time, length, and mass:
\begin{equation}
\label{eq:SSunits}
t_*\equiv\Omega^{-1}\,,\quad l_*\equiv\frac{\pi G\Sigma}{\Omega^2}\equiv \pi^{-1} L_T\,,\quad m_*\equiv\Sigma l_*^2\,.
\end{equation}

The length $l_*$ represents 
the pressure scale height when Toomre's stability indicator 
$Q_T\equiv\Omega\cs/(\pi G\Sigma)$ is unity, 
and characteristic sound speed $c_s\to l_*/t_*$.
Another important lengthscale is the ``Toomre length" $L_T$, 
which carries an additional factor of $\pi$, 
corresponding to the 2D Jeans length $c_s^2/(G\Sigma)$,
or half the wavelength of the marginally unstable mode, at $Q_T \sim 1$. 
Note that $\cs$ (defined in terms of the \emph{total} pressure) 
is not included among the control parameters: instead, 
$\cs$ needs to be regarded as an \emph{output} parameter of our models, 
determined by the balance between turbulent heating and radiative cooling, 
though again we expect the steady-state $\cs$ to correspond to $Q_T \sim 1$.

To a state determined by $(\Sigma,\Omega)$ and $Q_T\sim 1$ 
one must add physical parameters to describe the equation of state and cooling of the gas. 
In scale-free 2D \citep{Gammie2001} or 3D \citep{Shi2014} local simulations, 
there were only two such parameters: 
an adiabatic index $\gamma$ relating the gas internal energy per unit mass to its 2D or 3D gas pressure 
$U_{\rm gas}=P_{\rm gas}/(\gamma-1)$, and a Newtonian characteristic cooling timescale $t_{\rm cool}$, 
or equivalently, 
the dimensionless product $\tau_{\rm cool} = \Omega t_{\rm cool}$. 
Our case is somewhat more complex because we model the radiation field by which the matter cools, 
such that instantaneous cooling rates are determined by physical values of $\Omega,\Sigma$ and the variable $Q_T$ (or equivalently $\cs$), 
and we are particularly interested in regimes where the radiation also contributes significantly to the pressure support of matter.

\subsection{Relative Contribution from Radiation and Gas Pressure}

Before addressing cooling, 
we discuss the equation of state and temperature profiles, 
having in mind conditions near the midplane where optical depths are large and the gas and radiation temperatures are equal. 
From the dynamical units (\autoref{eq:SSunits}), one can form a combination with units of pressure: $P_*=m_* l_*^{-1} t_*^{-2} = \pi G\Sigma^2$, as well as units of density $\rho_*=m_* l_*^{-3}= \Omega^2/\pi G$.


At $Q_T\sim 1$, 
we can generally express the midplane or some characteristic total pressure and density as $P = f_p P_*$ and $\rho=f_\rho\rho_*$, 
where $f_p, f_\rho=\mathcal{O}(1)$ are dimensionless constants depending on the detailed vertical structure and $Q_T$, 
while $T$ is the midplane or some characteristic temperature.

The total pressure is the sum of the gas and radiation pressure:
\begin{align}\label{eq:quartic}
    \frac{\rho\kB T}{\mu} + \frac{aT^4}{3}=P_{\rm gas} + P_{\rm rad} &= P\,,\nonumber\\[1ex]
\mbox{or}\quad \frac{f_\rho\Omega^2}{\mu \pi G}(\kB T) + \frac{\pi^2}{45}\frac{(\kB T)^4}{(\hbar c)^{3}} &= f_p\pi G\Sigma^2\,.
\end{align}
Here, $\mu$ is the molecular weight of gas. This is directly analogous to the Eddington quartic for the central conditions in non-degenerate stars \citep[e.g.][]{GoodmanTan2004}.
As in that case, 
\autoref{eq:quartic}
can be regarded as a quartic equation for the temperature.
There is a single positive root to the quartic, whereby $T$ becomes a function of $(\Sigma, \Omega)$.

It is convenient to introduce a symbol for the ratio of pressures:
\begin{equation}\label{eq:prg-def}
    \prg \equiv\frac{P_{\rm rad}}{P_{\rm gas}} = \dfrac{aT ^3\mu}{3\rho k_{\rm B}}\,.
\end{equation}
This is equivalent to $(1-\beta)/\beta$ in the stellar-structure literature (e.g., \citealt{GoodmanTan2004}).
At $\prg=1$, the two terms on the left side of \autoref{eq:quartic} are equal, such that $(T,\Omega,\Sigma,\mu)$ are related at $\Pi=1$ by
\begin{align}\label{eq:Teq}
    \kB T &= \left(\frac{45 f_\rho\Omega^2}{\pi^3 G\mu}\right)^{1/3}\hbar c\,,\quad\mbox{and}\nonumber\\[1ex]
    f_p^3 f_\rho^{-4} &= \frac{360}{\pi^{2}}\frac{(\hbar c)^3}{(\pi G)^7}\frac{\Omega^8}{\mu^4\Sigma^6}\,,\\
    &\approx 6.4 \Omega_{-8}^{8}\Sigma_{5}^{-6}\mu_{0.6}^{-4}\,,\nonumber
\end{align}
with $\Omega_{-8}=\Omega/(10^{-8}\unit{rad\,s^{-1}})$, $\Sigma_5=\Sigma/(10^5\unit{g\,cm^{-2}})$, and $\mu_{0.6}=\mu/(0.6m_p)$, for proton mass $m_p$.

More generally, \autoref{eq:quartic} can be rewritten in terms of the pressure ratio as
\begin{equation}
\begin{aligned}
    \Pi^{1/3}+\Pi^{4/3} &= \left[f_p^3 f_\rho^{-4}\frac{\pi^{2}}{45}\frac{(\pi G)^7}{(\hbar c)^3}\mu^4\Sigma^6\Omega^{-8}\right]^{1/3}\\
    &\approx \left[1.3 f_p^3 f_\rho^{-4} \mu_{0.6}^4\Sigma_5^6\Omega_{-8}^{-8}\right]^{1/3},
\end{aligned}
\label{eq:quarticPi}
\end{equation}
the solution of which gives the function $\Pi(\Sigma, \Omega)$. Here it is evident that $\prg = {\rm Const.}$ solutions have $\Sigma \propto \Omega^{4/3}$. Furthermore, requiring that vertical hydrostatic equilibrium is satisfied constrains $f_p$ and $f_\rho$ as functions of $Q_T$ only.

\begin{figure}[htbp]
\centering
\includegraphics[width=0.5\textwidth,clip=true]{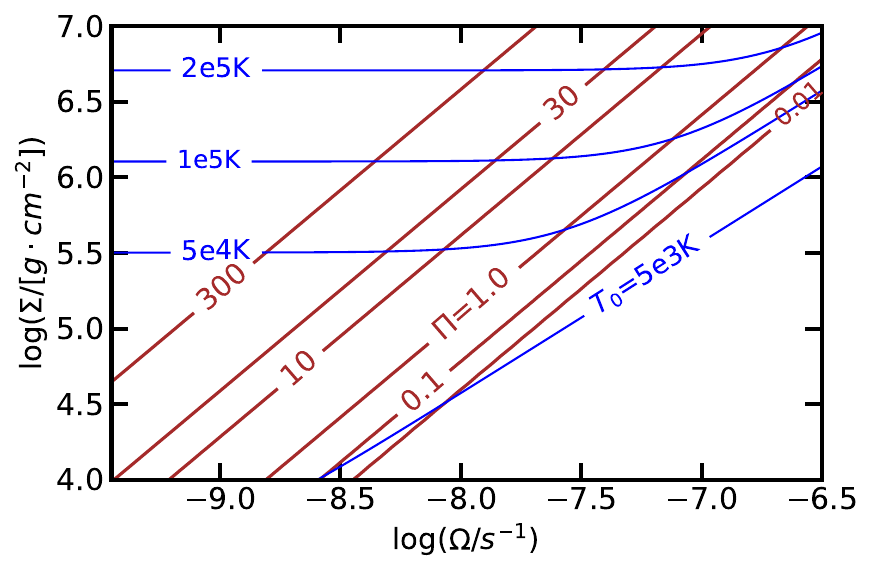}
\caption{Contours of pressure ratio $\Pi$ (brown red) and the midplane temperature $T_0$ (blue) in the $\Sigma, \Omega$ plane, assuming a vertically constant $\Pi$ and $Q_T=1$ initial condition as described in \S \ref{sec:initial}. $Q_T$ is defined formally in \autoref{eqn:QTinitial}.}
\label{fig:contours_Pi_T}
\end{figure}

As a quantitative example, 
\autoref{fig:contours_Pi_T} shows contours of the   pressure ratio $\Pi$ and the midplane temperature $T_0$ (subscript ``0" specifically denotes midplane) as functions of $(\Sigma, \Omega)$, 
for a specific self-similar vertical disk profile that assumes an altitude-independent $\prg$ and a mass-weighted $Q_T=1$ (see \autoref{sec:initial} for details; $Q_T$ is defined formally in  \autoref{eqn:QTinitial}). 
This family of solutions of  \autoref{eq:quartic} and \autoref{eq:quarticPi} holds specifically for $f_p(Q_T=1) = 0.66$ and $f_\rho(Q_T=1) = 0.74$.

Neglecting all constants including $f_p$ and $f_\rho$, it is easy to see that 
\begin{equation}
\begin{aligned}
T &\propto \Sigma^2\Omega^{-2}, &\Pi &\propto \Sigma^6 \Omega^{-8} &\mbox{if }\Pi &\ll 1   \\
T &\propto \Sigma^{1/2}, &\Pi &\propto \Sigma^{3/2} \Omega^{-2} \quad\quad &\mbox{if }\Pi &\gg 1.
\end{aligned}
\label{eqn:T_prop}
\end{equation}
One can understand the situation more easily by comparing these scalings with the change in slope of the contours in \autoref{fig:contours_Pi_T} across $\Pi = 1.0$. 
From the gas to radiation pressure dominated regime, the contours of $\Pi$ become sparser and the $T_0$ contours flatten out after crossing $\Pi=1.0$.

To summarize, instead of considering only the gas pressure and internal energy in the equation of state, the total internal energy is given by the sum of gas and radiation intermal energy $U = P_{\rm gas}/(\gamma-1) + 3P_{\rm rad}$, 
with their relative contribution determined by $\Pi$ as a function of $(\Sigma, \Omega)$, which itself follows different scalings in the limits $\Pi\gg1$ and $\Pi\ll1$.
In the remainder of this paper, we take $\gamma=5/3$ as for a fully ionized plasma.

\subsection{Cooling Timescales}

Although not explicitly involved in the pressure ratio $\Pi$, the opacity ($\kappa$) should be important for the \emph{strength} of the turbulence, because the heating rate and cooling rate must balance on average in a gravito-turbulent state, with the latter being sensitive to $\kappa$ through the vertical optical depth of the system $\tau$. Nevertheless, with $\kappa$ generally being a function of $(\rho, T)$, the cooling rate is also expressible in terms of the combination $(\Sigma, \Omega)$ with the constraint $Q_T\sim 1$. Specifically, for constant opacity applicable to high temperature, we have $\tau \equiv\kappa\Sigma/2$, 
and the radiative energy flux from each of the two disk surfaces can be approximated as \citep{Johnson2003}

\begin{equation}
    F_{z,\rm max} = \sigma T_{\rm eff}^4 = \dfrac{8 \sigma T_0^4}{3 \tau} = \dfrac{16 \sigma T_0^4}{3\kappa \Sigma}.
    \label{eq:Fzmax}
\end{equation}

One can then calculate the realistic cooling timescale 
\begin{equation}
t_{\rm cool} = U_{2D}/2F_{z,\rm max},
\label{eqn:cooling_time}
\end{equation}
which is the inverse of cooling rate. Here $U_{\rm 2D} = \int_z U \mathrm{d}z = f_{U}  P_*l_*$ is the vertically integrated internal energy per unit area, while $f_{U}$ is yet another order-unity coefficient determined by the vertical profile, slightly more complicated than $f_\rho, f_p$ in the sense that it has a small dependency on the pressure weighting $\Pi$ that manifests mainly in the $\Pi \sim 1$ transition region. The dimensionless $\tau_{\rm cool}= \Omega t_{\rm cool}$ is the cooling time measured in units $t_* = \Omega^{-1}$. 

\begin{figure}[htbp]
\centering
\includegraphics[width=0.5\textwidth,clip=true]{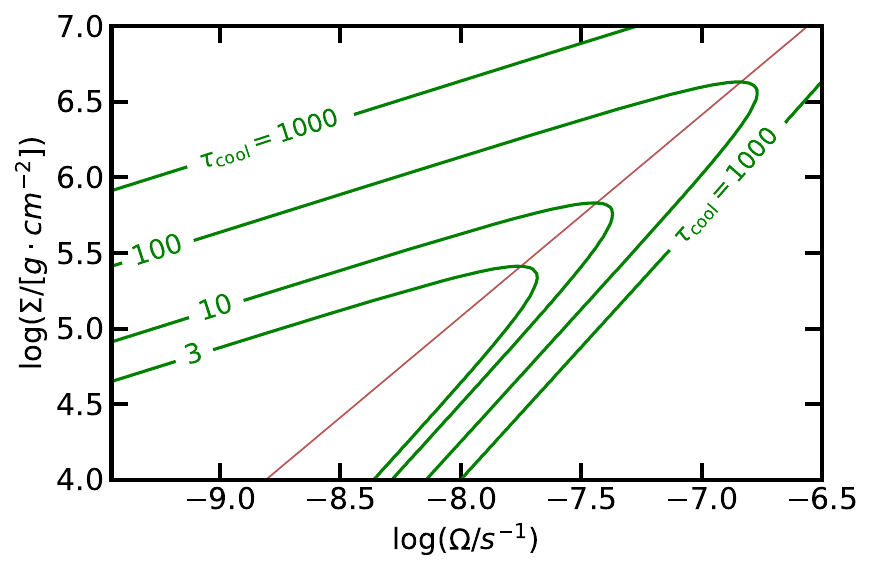}
\caption{Contours of $\tau_{\rm cool}$ in the $\Sigma, \Omega$ plane, assuming a vertically constant $\Pi$ and $Q_T=1$ initial condition as described in \S \ref{sec:initial}. The brown line is the $\Pi=1$ contour.}
\label{fig:contours_tcool}
\end{figure}

Similar to \autoref{fig:contours_Pi_T}, we plot contours of $\tau_{\rm cool}$ for our specific example of vertical distribution in Figure \ref{fig:contours_tcool}, which follow from numerical results of  $T(\Sigma, \Omega)$ (\autoref{fig:contours_Pi_T}) and the specific choice for $f_U(Q_T=1) = 1-1/(2+2\prg)$ corresponding to our self-similar vertical profile. 
The trends here can also be understood in a simple manner by checking proportionality scalings in two different regimes combining \autoref{eqn:T_prop} and \autoref{eqn:cooling_time}. 
When $\kappa$ is constant, $F_z \propto T^4\Sigma^{-1}$, applying $U_{2D} \sim  P_* l_*$ we have 
\begin{equation}
\begin{aligned}\label{eq:tau_cool_scale}
\tau_{\rm cool} &\propto \Sigma^{-4} \Omega^{7} &\mbox{if } \Pi \ll 1   \\
\tau_{\rm cool} &\propto \Sigma^2 \Omega^{-1} &\mbox{if } \Pi \gg 1.
\end{aligned}
\end{equation}

The scalings in \autoref{eq:tau_cool_scale} imply that as we start from the lower-right gas pressure dominated regime, 
for increasing $\Sigma$ or decreasing $\Omega$, $\tau_{\rm cool}$ always reaches a minima (maxima for cooling rate) 
around $\prg \approx 1 $ before increasing again in the radiation dominated regime. 

Specially, we can write down a more explicit expression for the cooling timescale in the $\prg \gg 1$ regime, where $a T_0^4 = 3 P_0 = 3f_p P_*$:   
\begin{equation}
\begin{aligned}
    \tau_{\rm cool} &\approx \dfrac{ f_U P_* l_* \Omega \times 3 \kappa \Sigma}{8 c a T_0^4 }  = \dfrac{  f_U c_s } {8 f_p c } \times \kappa \Sigma \\&= \dfrac{ \pi f_U\kappa G\Sigma^2} {8 f_p c\Omega }, \quad \Pi \gg 1
    \end{aligned}
\label{eqn:t_cool_radiation}
\end{equation}
In \S \ref{sec:rad_frag_results} we will see this is closely related to a radiative diffusion criterion. 

It has been shown by extensive simulations 
that in the classical gas pressure dominated regime, 
the fragmentation boundary is roughly determined by an approximate value of $\tau_{\rm cool} \sim 3$ \citep{Gammie2001,Johnson2003,Rice2003,Shi2014}. 
Another way to interpret this conclusion is that in thermal equilibrium, 
turbulent heating must balance cooling such that $\alpha \sim \tau_{\rm cool}^{-1}$, 
and the critical $\alpha$ a steady state gas disk could allow cannot exceed $\sim$ 0.3. 
Such a boundary would suggest that the lower-left region of the parameter space within $\tau_{\rm cool} < 3$ 
represent transient states that will fragment and cannot be maintained by gravito-turbulence, 
unless other heating sources are involved. 

Following \citet{Jiang11}, we expect the disk to be more subject to fragmentation when radiation pressure dominates.
Specifically, 
the maximum $\alpha$ that a quasi-steady state could support decreases with the growth of $\Pi$. As a result, 
the fragmentation regime can no longer be constrained by some universal value of $\tau_{\rm cool}$, 
but should span to larger $\tau_{\rm cool}$ for increasing $\Pi$, 
possibly covering any value of $\tau_{\rm cool}$ when $\Pi \gg 1$.

Realistic opacities can have complicated dependencies at low temperature, 
which should modify the cooling rate contours in the gas-dominated regime \citep[e.g.][Figure 7]{Johnson2003}, 
especially when temperature drops to around 2000K
and opacity becomes extremely small \citep{TQM}. In this paper, however, we adopt a constant opacity in order to isolate the effects of the pressure ratio $\Pi$.

\subsection{Accretion Rates in Steady-state}

To link local simulations with global structure of AGN disks, 
we introduce the scaling for a kind of ``expected" local accretion rate $\dot{M}$. 
Under assumption that in steady-state where heating provided by gravito-turbulence balances radiative cooling rates given by $\tau_{\rm cool}$, 
every point on the $(\Sigma,\Omega)$ plane also corresponds to a local accretion rate \citep{Pringle1981,Jiang11}: 

\begin{equation}
    \dot{M} = \dfrac{8\pi F_{z,\rm max}}{3\Omega^2}\propto F_{z,\rm max}\Omega^{-2},
    \label{eqn:dot_M}
\end{equation}

\begin{figure}[htbp]
\centering
\includegraphics[width=0.5\textwidth,clip=true]{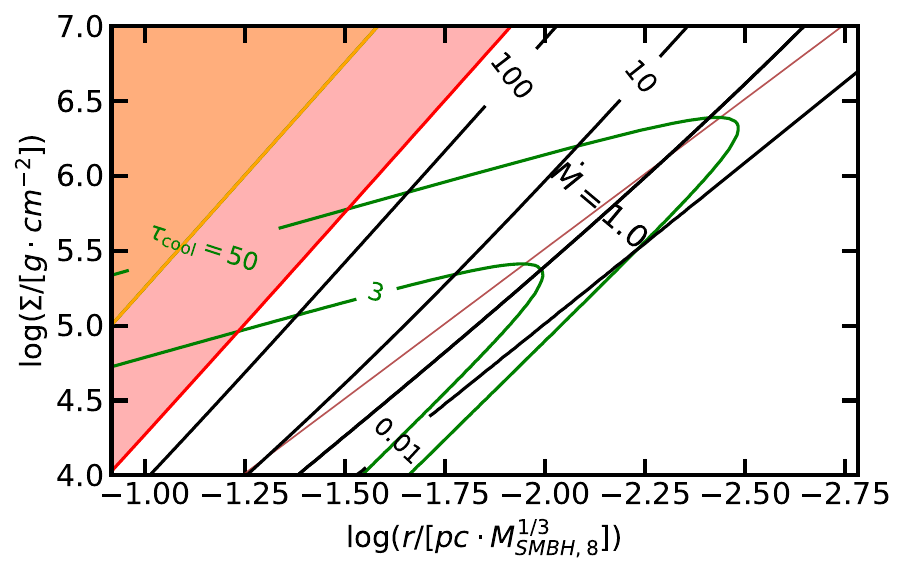}
\caption{Contours of $\dot{M}$  ({\it black, in $\unit{M_\odot\,yr^{-1}}$}) and $\tau_{\rm cool}$ ({\it green}), assuming the initial condition as described in \S \ref{sec:initial}, and turbulent heating that balances the cooling rate given in Figure \ref{fig:contours_tcool}.
The brown line is the $\Pi=1$ contour.
The abscissa is re-written in terms of radius $r$  scaled by the SMBH mass, where $M_{\textsc{smbh},8}=M_{\textsc{smbh}}/10^8 M_{\odot}$. 
We also map out highly super-Eddington regions where $\dot{M}$ exceeds 100 times the Eddington value for $M_{\textsc{SMBH}} = 10^8$ (pink region) and $10^9 M_\odot$ (orange region), for radiative efficiency $\eta =0.1$ and Eddington luminosity.}
\label{fig:contours_Mdot}
\end{figure}

The proportionality can also be derived from
\begin{equation}
     \dot{M} \propto \alpha c_s^2 \Sigma \Omega^{-1}, \alpha \sim \tau_{\rm cool}^{-1}, 
\end{equation}
at a $Q_T= 1$ steady-state \citep{Gammie2001}. For our specific vertical profile, 
we plot in \autoref{fig:contours_Mdot} exact contours of $\dot{M}$ calculated from \autoref{eqn:dot_M} 
in our numerical profiles. 
Combining \autoref{eq:Fzmax} and \autoref{eqn:dot_M}, in two limits $\dot{M}$ conforms with the scalings
\begin{equation}
    \begin{aligned}
        \dot{M} & \propto \Sigma^7\Omega^{-10}&\mbox{if } \Pi \ll 1\\
        \dot{M} & \propto \Sigma\Omega^{-2} &\mbox{if } \Pi \gg 1.
    \end{aligned}
\end{equation}

As we follow contours of constant $\dot M$ in \autoref{fig:contours_Mdot}, 
we see how $\tau_{\rm cool}$ and $\Pi$ varies radially along a disk with given accretion rate. Because $\dot{M}$ contours are always steeper than $\Pi$ contours, 
a $Q_T \sim 1$ accretion disk always becomes radiation dominated at sufficiently small $r$.  For example,
the $\dot{M} =1 M_\odot $/yr contour (roughly half the Eddington rate for $M_{\rm SMBH}=10^8 M_\odot$) intersects $\prg = 1$ 
a bit below 
$\tau_{\rm cool}\sim 50$. 
In fact, as $r$ increases, both $\Pi$ and $\tau_{\rm cool}$ decrease along each $\dot{M}$ contour, albeit with the latter being much more sensitive. Beyond some large radius, $\tau_{\rm cool}$ falls below a critical value which is either $\sim 3$ for $\Pi \ll 1$ or possibly larger for $\Pi > 1$, 
and we inevitably enter a star-forming region of fragmentation, which may also provide self-regulated heating to maintain $Q_T\sim 1$ \citep{TQM}.
Since star formation would therefore be able to  strip away large fractions of $\dot{M}$ before reaching the inner disk, 
this suggests it would be difficult 
for a super-Eddington constant-$\dot{M}$ flow to penetrate into inner gravito-turbulent parts of the disk. 
We map out regions where $\dot{M}$ exceeds 100 Eddington value for $M_{\rm SMBH} = 10^8$ (pink region) and $10^9 M_\odot$ (orange region), assuming accretion efficiency $\eta =L/\dot{M}c^2 = 0.1$ 
and the luminosity $L$ equals the Eddington luminosity. 
Such accretion rates may be achieved temporarily if mass were dumped into the disk by violent events such as mergers or Tidal Disruption Events (TDEs), 
or if the formed massive stars themselves can lose mass that feeds the central SMBH \citep{Cantiello2021}. 

Apart from the outer boundary of the gravito-turbulent region 
constrained by some lower limit of $\tau_{\rm cool}$, 
in realistic situations we also need to consider an inner boundary of the $Q_T=1$ disk region given by the upper limit $\tau_{\rm cool} \lesssim  50$.
For larger $\tau_{\rm cool}$,
the heating rate of MRI turbulence, 
which typically gives $\alpha_{\rm MRI} \sim 0.02$ \citep{Beckwith2011,Simon2012}, 
is already adequate to heat the disk up to $Q_T>1$ and shut off gravitational instability, 
and we enter the inner standard thin disk region \citep{Shakura1973}. 
In other words, 
even if a $Q_T\sim 1$ disk heated solely by gravito-turbulence were stable at sufficiently long $\tau_{\rm cool} \gg 3$ in the far upper left region of \autoref{fig:contours_Mdot}, 
solutions in that part of parameter space would not be physical due to extra heating from MRI that would realistically be present.

It follows from the above discussion that in the $(\Sigma, \Omega)$ parameter space, 
a classical model for a $Q_T\sim 1$, 
sub-Eddington gas pressure dominated disk can be simply parameterized as a low-$\dot{M}$ contour constrained within $\tau_{\rm cool} \in [3, 50]$, 
representing a gravito-turbulent region sandwiched between an inner gravitationally stable standard thin disk powered by MRI, 
and an outer star-forming region \citet{Goodman2003,TQM}. 
While the above describes behavior at low accretion rates, uncertainty 
in the fragmentation limit for higher $\dot{M} \gtrsim M_\odot$/yr disks in the $\prg \gtrsim 1$ regime prevents us to assert the same conclusion for the radiation dominated regime. 
In this paper, 
we will apply hydrodynamic simulations with full radiative transfer to check the validity of the classical fragmentation condition in the gas pressure dominated regime, 
and to explore how the fragmentation condition is modified at increasing values of $\prg$. 

\section{Methods}
\label{sec:setup}

\subsection{Equations Solved}

We adopt the 3D shearing box configuration in \texttt{Athena++} \citep{Stone2020}, 
and solve ideal hydrodynamic equations coupled with the time-dependent, 
frequency-integrated 
radiation transport equation for specific intensities over discrete angles \citep{Jiang2014,Jiang2021,Goldberg2021}:

\begin{equation}
    \frac{\partial \rho}{\partial t}+\boldsymbol{\nabla} \cdot(\rho \mathbf{v})=0
\end{equation}

\begin{equation}
\begin{aligned}
    \frac{\partial(\rho \mathbf{v})}{\partial t}+\boldsymbol{\nabla} \cdot\left(\rho \mathbf{v} \mathbf{v}+\mathbf{P}_{\mathrm{gas}}\right)=-\rho \boldsymbol{\nabla} \Phi
    \\
    -2 \rho \Omega \hat{\mathbf{z}} \times \mathbf{v}+2 q \rho \Omega^{2} x \hat{\mathbf{x}}-\rho \Omega^{2} z \hat{\mathbf{z}}-\mathbf{G}_{r}
    \end{aligned}
\end{equation}

\begin{equation}
    \begin{aligned}
\frac{\partial E}{\partial t} &+\boldsymbol{\nabla} \cdot(E+P_{\rm gas}) \mathbf{v}=-\rho \mathbf{v} \cdot \boldsymbol{\nabla} \Phi \\
&+\rho \Omega^{2} \mathbf{v} \cdot(2 q x \hat{\mathbf{x}}-z \hat{\mathbf{z}})-cG^0_r
\end{aligned}
\end{equation}
\begin{equation}
    \frac{\partial I}{\partial t}+c \mathbf{n} \cdot \boldsymbol{\nabla} I=S(I, \mathbf{n})
\end{equation}

In these equations, $\rho$ is the gas density, $\mathbf{v}$ is the 3D flow velocity and $q=3/2$ is the Keplerian shear parameter. $\mathbf{P}_{\mathrm{gas}}$ and $P_{\rm gas}$ are the gas pressure in tensorial and scalar form, respectively: i.e., $\mathbf{P}_{\rm gas}=P_{\rm gas}\mathbf{1}$ if $\mathbf{1}$ represents the unit tensor. The gas mean molecular weight $\mu = 0.6m_p$ is that 
of a fully ionized gas at solar abundance. $E=U_{\rm gas}+\rho v^{2} / 2$ is the sum of gas internal energy $U_{\rm gas}=P_{\rm gas}/(\gamma-1)$ and the kinetic energy $\rho v^2/2$ where $\gamma=5/3$. 
The vertical component of the external gravity by the central star is added as source terms in the equation for momentum and energy transport.
The source terms $\mathbf{G}_{r}$ and $G^0_r$ are the time-like and space-like components of the radiation four-force \citep{Mihilas1984}. $I$ is the frequency-integrated intensity
and $\mathbf{n}$ is the photon propagation direction unit vector. 

The hydrodynamic equations are solved using the standard Godunov method in \texttt{Athena++} \citep{Stone2020}. We use the second-order Van-Leer method for the time integration, and the HLLC Riemann solver to calculate the flux for hydrodynamic quantities. We adopt second-order reconstruction for intensity as well as hydrodynamic quantities. The disk self-gravitational potential is obtained by  solving the Poisson equation
\begin{equation}
    \boldsymbol{\nabla}^{2} \Phi=4 \pi G \rho,
\end{equation}
using fast Fourier transforms \citep{Koyama2009,Kim2011}.
 For each radiative transfer calculation, we update the intensity in the co-moving frame $I_0(\mathbf{n}_0)$ 
(which is Lorentz-transformed from the lab frame $I(\mathbf{n})$) by the source term $S_0$ in the comoving frame:
\begin{equation}
S_{0}\left(I_{0}, \boldsymbol{n}_{0}\right)=c \rho \kappa_{P}\left(\frac{c a T^{4}}{4 \pi}-J_{0}\right)+c \rho\left(\kappa_{s}+\kappa_{ R}\right)\left(J_{0}-I_{0}\right)
\end{equation}
then convert it back to the lab frame. 
Here $\kappa_s$ is the electron scattering opacity, $\kappa_R$ is the Rossland mean opacity, and $\kappa_P$ is the Planck opacity. 
In this work, we adopt a constant electron scattering opacity $\kappa_s = 0.4\text{cm}^2/$g and a fiducial $\kappa_R =0.05\text{cm}^2/$g that is non-zero but subdominant. 
We also choose $\kappa_P= 0.05\text{cm}^2/$g under the gray opacity approximation. 
While realistic Planck opacities may be larger \citep{Jiang2020}, 
this value suffices as long as the radiation field is adjusted to be in temperature equilibrium with gas, 
which turns out to be the case in our simulations. See \citet{Jiang2021} for more details on implementation of radiation. 

\subsection{Initial Conditions}
\label{sec:initial}

The initial equilibrium profile of our self-gravitating disk is assumed to be horizontally homogeneous, 
while the vertical distribution is derived semi-analytically. 
The initial surface density $\Sigma$ and orbital frequency $\Omega$ are needed as input parameters. 
For convenience, 
we define a midplane gravitational instability factor that is a proxy for Toomre $Q_T$:
\begin{equation}\label{eq:altQ}
Q = \dfrac{\Omega^2}{2 \pi G \rho_0},
\end{equation}
where $\rho_0$ is the density in the midplane (subscript denotes midplane quantities). 
Following the method of \citet{Jiang11}, 
with a given $Q$ we derive density and pressure distribution $\rho(z)$ and $P(z)$ for a self-gravitating constant-$\Pi$ polytrope within the photosphere, 
combined with an isothermal radiation field outside the photosphere; 
see Appendix~\ref{appendix} for details. 
We confirm that initial hydrostatic equilibrium holds before the disk starts to cool.

From the expressions for midplane $\rho_0$ and pressure $P_0$, 
we find (see \autoref{eq:rho0_P0}) the coefficients $f_\rho(Q)$ and $f_p(Q)$ for our specific version of the midplane EoS quartic (\autoref{eq:quartic} \& \autoref{eq:quarticPi}) that determines the solution of $\prg$, $T_0$. 
The problem remaining is to choose appropriate values of $Q$ and to connect with the classical Toomre value $Q_T$, technically defined in a razor-thin 2D disk \citep[e.g.][]{Johnson2003}. The sound speed, written explicitly as
\begin{equation}
    \cs = \left(\dfrac{P_{\rm gas}+P_{\rm rad}}{\rho}\right)^{1/2}
\end{equation}
is well defined in an isothermal 3D disk as a constant, 
with $Q_T = \sqrt{2/\pi} Q$. 

When $\cs$ is not a constant, 
as in our polytropic vertical profile, it is still possible to define some $Q_T$ which reflects vertically-averaged properties, such as 
\begin{equation}
    Q_T = \dfrac{[ c_s^2]_{\rho}^{1/2}\Omega}{\pi G\Sigma} 
    \label{eqn:QTinitial}
\end{equation}
where $[c_s^2]_{\rho}$ is a density-weighted average of the square sound speed, 
directly connected to the vertically integrated pressure:
\begin{equation}
      [c_s^2]_{\rho} \equiv \dfrac{\int_z c_s^2 \rho dz}{\int_z \rho dz}  = \dfrac{\int_z (P_{\rm gas}+P_{\rm rad}) dz}{\Sigma}
\end{equation}

This $Q_T$, defined similarly to Equation 9 of \citet{Booth2019} or Equation 13 of \citet{2017MNRAS.471..317R}, 
has the advantage of capturing the vertically averaged properties and will smoothly connect to the global average quantity $Q_T$ in \S \ref{sec:diagnostic}.
Combined with \autoref{eqn:P2D}  expressing $P_{\rm 2D} = \int_z P dz$ in terms of $Q, \Sigma, \Omega$, 
we see that the midplane $Q$ and the average $Q_T$ are linked through
\begin{equation}
    Q_T = \sqrt{\frac{ Q I_{4}(Q)}{16\left[I_{3}(Q)\right]^{3}} },
\end{equation}
Through approximations given in \autoref{eq:structure_approximations}, we understand that $Q_T \propto Q$ when they are large but $Q_T \propto \sqrt{Q}$ when $Q\ll 1$ (the strongly self-gravitating limit).


Referring to \autoref{eqn:U2D}, the coefficient $f_U$ can be more simply expressed as:
\begin{equation}
    f_U = Q_T^2 \left[1-\dfrac{1}{2(\Pi + 1 )}\right]
\end{equation}

With a ``reference plane" parameterized by $Q_T=1$, corresponding to $Q=0.76$, we solve quartic equations for $\prg$ and $T_0$ as functions of $\Sigma, \Omega$ as shown in \autoref{fig:contours_Pi_T}, making use of $\rho_0 = f_\rho \rho_* = 0.74 \rho_*$ and $P_0 = f_{P}P_* = 0.66 P_*$.
Combining the vertically integrated internal energy density per area $U_{\rm 2D}$ with the vertical radiative flux beyond the photosphere $F_{z,\rm max}$, 
we can calculate $\tau_{\rm cool}(\Sigma, \Omega)$ (\autoref{eqn:cooling_time}) as shown in \autoref{fig:contours_tcool}. 
Lastly, 
we can link the local parameter space with global accretion rates using $\dot{M}(\Sigma, \Omega)$ (\autoref{eqn:dot_M}) as shown in \autoref{fig:contours_Mdot}. 
One can refer back to \S \ref{sec:formulation} for general scalings of these contours in the $\Pi \gg 1$ and $\Pi \ll 1$ limits 
\footnote{The scalings are valid for any polytropic vertical density profile. 
Numerically, 
the initial total vertically integrated $U_{2D}$ does not follow the analytical expectation $f_U P_* l_*$ exactly due to the existence of an isothermal radiation field outside the photosphere, 
but the deviation is very small.}.

Practically, 
in our simulations, 
by default we start from a $Q_{T} \approx 1.05, Q \approx 0.82$ state slightly hotter than in the reference plane, 
such that we allow gravitational instability to gradually develop during an initial passive cooling phase. 
As a result, contours in \autoref{fig:contours_Pi_T}, \autoref{fig:contours_tcool} are not exactly the initial $\prg$ and $\tau_{\rm cool}$ of our profiles, 
but rather the  ``expected" values of some time-average value for $ \tau_{\rm cool}$ and $\prg $ \textit{if} turbulent heating can support a steady state at $ Q_T \approx 1$ with a similar vertical structure as the initial condition. 

As we shall see, 
our vertical structure is a useful approximation for the final gravito-turbulent quasi-steady states within the photosphere, at least for gas pressure dominated cases, 
but usually on average $Q_T \gtrsim  1.1$ in the final states, 
which results in a generally hotter state with larger time-averaged $\prg'$ than the ``expected" values. 
In fragmentation cases, 
there are no steady-state values to be measured so the reference contours are not self-consistent with the outcome, 
but if additional heat source is considered the disk may still settle into a steady state \citep{TQM}. 


{We adopt code units such that $\Omega$, $\Sigma$, 
and the Toomre length
$L_T\equiv\pi^2 G\Sigma/\Omega^2$ are unity; our length and mass units are therefore larger than the quantities $l_*$ and $m_*$ defined in \autoref{eq:SSunits} by factors of $\pi$ and $\pi^2$, respectively.
Expressed in such units, 
we have a universal initial profile for $\rho, T$ parametrized only by $Q$ regardless of $\Sigma, \Omega$ (e.g. \autoref{fig:verticalall}).
The code unit for energy density is therefore $\Sigma \Omega^2 L_T = \pi^2 G\Sigma^2$, 
and that of temperature is $(L_T\Omega)^2 $.}

We initialize a decaying turbulence field using the default setup of \texttt{Athena++}, which distributes an assigned total turbulence kinetic energy $E_{turb}$ across wavenumber $m_{\rm min}=1$ to $m_{\rm max}=16$ with a spectral slope that we choose to be -2. 
We choose $E_{turb}/U \sim \tau_{\rm cool}^{-1}$ such that $E_{turb}$ may smoothly connect to an instability-generated turbulence that balances cooling. 
In a long-term steady state, 
the outcome should not depend on details of the initial turbulence field. 
We also tested some representative fragmentation cases with different $m_{\rm max}$ and confirmed convergence.

\subsection{Box Sizes and Resolution}

The default box size for our simulations is $L_x\times L_y\times L_z = 8\times 8\times 2 (L_T^3)$. 
Due to the computational expense of full three-dimensional RHD, 
the default low resolution with which we run long-term simulations for quasi-steady or marginally stable cases is $N_x\times N_y\times N_z = 128\times 128\times 32$. 
In certain cases with high optical depth, 
we also found it necessary to simulate with a larger box size to prevent strong outflow, such that 
$L_z = 4$ and $N_z =64$. 
For most of our short-term fragmenting cases, we can afford to test with a doubled resolution $N_x\times N_y\times N_z = 256\times 256\times 64$ keeping the default box size. 
In certain radiation dominated cases, where it is necessary to resolve the Jeans length defined by gas pressure rather than total pressure, 
we also performed extra tests with smaller box sizes and higher resolution; see \S \ref{sec:rad_frag_results} for details. \autoref{tab:para} lists the full set of input parameters for all runs discussed in the paper.

\subsection{Boundary Conditions}
For radiation and hydrodynamic variables we apply the standard shearing-periodic boundary condition in $x$ and periodic boundary condition in $y$. 
We implement an open outflow boundary condition in $z$ by setting the density, pressure and radiation intensity in the boundary cells to the same values as the last active cells. 
Additionally, 
we copied both the velocity and the radiative flux from the final cell to the ghost zones in the $z$ direction, 
but reset the velocity and radiative flux 
in the $z$ direction 
to zero if flows are directed into the simulation box to prevent artificial mass and energy injection.  
The Poisson solver for $\Phi$ applies shearing-periodic boundary condition in $x$ periodic boundary condition in $y$, 
and vacuum boundary conditions in $z$. 
We implement floor values such that $T$ does not fall below $10^{-3}$ of the initial midplane temperature $T_0$, 
and $\rho$ does not fall below $10^{-6}$ of the initial midplane density $\rho_0$.
Because our floor values are sufficiently small, 
in most of our simulations we were able to prevent significant mass loss and achieve mass conservation without the need of artificial mass-rescaling, e.g. applied in \citet{Booth2019}.

\subsection{Diagnostics}
\label{sec:diagnostic}
To facilitate analysis of our simulation results, 
we first introduce notation for some scalar history variables obtained by averaging physical quantities over space and time. 
We define the volume average $\langle X\rangle$ as:
\begin{equation}
    \langle X\rangle \equiv \frac{\int X d x d y d z}{\int d x d y d z}.
\end{equation}
With this definition, instantaneous values of the dimensionless stress parameters that characterize angular momentum transport are defined as
\begin{equation}
\begin{aligned}
    \alpha_R &= \dfrac{\langle R_{xy} \rangle}{\langle P_{rad} + P_{gas}\rangle}, &\alpha_g = \dfrac{\langle G_{xy} \rangle}{\langle P_{rad} + P_{gas}\rangle}\\ R_{xy} &=  \rho v_x \delta v_y, &G_{xy} =  g_x g_y/4\pi G  
    \end{aligned}
\end{equation}
where $\delta v_y=v_y + q\Omega x$, the perturbed $y$ velocity subtracting out the background shear flow. {We have verified that contribution of the stress term from radiation viscosity \citep[e.g.][Equation 15]{Blaes2011} is negligible in our simulations, in contrast to the situation in low optical depth disk surface regions within coronae caused by magnetic dissipation and vertical temperature inversion  \citep{Jiangcoronae}. }

We define the density-weighted mean square sound speed
\begin{equation}
    \langle c_s^2\rangle_{\rho} \equiv \dfrac{\int_z c_s^2 \rho dx dy dz}{\int_z \rho  dx dy dz},
\end{equation}
and the midplane average of density
\begin{equation}
    \langle \rho \rangle_{\rm mid} \equiv \frac{\int  \rho(z=0) d x d y }{\int d x d y},
\end{equation}
such that the ``Toomre-like" history variables
\begin{equation}
     Q_{T} =\frac{\langle c_{s}^{2}\rangle_{\rho}^{1 / 2} \Omega}{\pi G \langle \Sigma\rangle}, Q = \dfrac{\Omega^2}{2\pi G \langle \rho \rangle_{\rm mid}},
\end{equation}
start exactly from initial values of $Q_{T}, Q$ at $t=0$ as defined in the \S \ref{sec:initial}. 

In the analysis of vertical profiles, 
we calculate the horizontally-averaged vertical distribution of quantity $X$ as a function of $z$ via:
\begin{equation}
    \langle X(z)\rangle\equiv \frac{\int X(x,y,z) d x d y}{\int d x d y}.
\end{equation}
To normalize the vertical coordinate,
we define a fiducial scale height as 
\begin{equation}
    h = \dfrac{\langle\Sigma\rangle }{2\langle \rho \rangle_{\rm mid}}.
\end{equation}

To achieve good convergence in the averaging of radiation to gas pressure ratio, we find that it is preferable to  define 
\begin{equation}
     \prg' \equiv \frac{\int P_{\rm rad} d x d y  dz}{\int P_{\rm gas} d x d y dz},\    \prg'(z) \equiv \frac{\int P_{\rm rad} (x,y,z)d x d y}{\int P_{\rm gas} (x,y,z)d x d y},
     \label{eqn:def_prg}
\end{equation}
instead of directly averaging $\Pi$.

The dimensionless cooling time is defined as 
\begin{equation}
    \tau_{\rm cool} = \dfrac{\int U dxdydz}{\int \sum |F_{z}|(z = \pm z_{\rm max}) dxdy}\Omega
\end{equation}
where $U$ is the internal energy density of gas and radiation, 
and $\int |F_z| (z = \pm z_{\rm max}) dxdy$ is the total energy cooling rate 
by radiative flux directed outwards from the two vertical boundaries.

Time averages of our history variable are denoted as
\begin{equation}
    \langle X\rangle_{t} \equiv \frac{\int X d t}{\int d t}
\end{equation}
where the integration is over the time after saturation, as indicated in \autoref{tab:para_turb}.

We also define the running time average of history variables starting from $t_i$ as 
\begin{equation}
    \langle X\rangle_{t'<t} (t) \equiv \frac{\int_{t_i}^{t} X d t'}{\int_{t_i}^{t} d t'};
\end{equation}
this is helpful in 
visualizing convergence of strongly fluctuating variables such as stress parameters.

\subsection{Determining bound objects}
\label{sec:bound}
To confirm fragmentation, 
we verify formation of gravitationally bound objects using the method of \citet{Mao2020}, which is an extension  of the GRID-core algorithm \citep{GRID}. 
In short, 
for each local minimum in the gravitational potential field $\Phi$, 
we first identify the closed contour with the largest value of $\Phi =\Phi_{\rm max}$ which contains no other local mimima. 
Within this region of interest (referred to as an HBP for ``hierarchical binding parent''), we narrow the region down to cells where
\begin{equation}
    \int E_{\rm turb}+ U + \rho (\Phi-\Phi_{\rm max}) < 0,
\end{equation}
is satisfied when integrated up to some contour $\Phi' \leq \Phi_{\rm max}$, meaning gas is bound relative to the potential contour $\Phi_{\rm max}$.  
Here $E_{\rm turb}$ is the turbulent kinetic energy relative to the center of mass of the region, which is referred to as an HBR (for ``hierarchical bound region''). 
With this definition, material within the HBR, as a whole, 
does not have enough turbulent kinetic and thermal energy to move into a neighboring potential minimum.  
If the gravitational well of an HBP is deep enough, 
the entire HBP becomes an HBR.

\begin{table*}
\centering
  \begin{tabular}{c|ccccccccc}
     \hline\hline
     Model & $\Sigma$(g/cm$^2$) & $\Omega$(s$^{-1}$) & $\tau_{\rm cool}(t=0)$ & $\Pi(t=0)$ &  Box Size $(L_T^3)$ & Resolution  & $ t_{\rm final} \Omega$ & Outcome\\
     \hline
     S2e4O7e-9 & $2.0\times 10^4$ & $7.0\times 10^{-9} $& 3.95 & 0.0042 & $8 \times 8\times 2$ & $128\times 128\times 32$ & 100  & Turbulence, Figure \ref{fig:verticalall}\\
     
     fiducial& $5.0\times 10^4$ & $1.2\times 10^{-8} $ & 4.56 & 0.013 & -- & --& 150  & --, Figure \ref{fig:fiducialsnapshot} - \ref{fig:verticalall}, \ref{fig:spectrum_comparison}   \\
      
     fiducial\_res&-- &--&--&--&-- & $256\times 256\times 64$  & 100  & --, Figure \ref{fig:alpharunningaverage},\ref{fig:verticalall} \\
     
     S1e5O2.2e-8 & $1.0\times 10^5$&$2.2\times 10^{-8} $  & 19.32 &0.0068 & --&$128\times 128\times 32$ & 150   & --, Figure \ref{fig:verticalall}  \\
     
     S2e5O3.6e-8  & $2.0\times 10^5$ & $3.6\times 10^{-8} $  & 38.19 & 0.0084 & -- & --  & 250 & --, Figure \ref{fig:verticalall}  \\
     
     S5e5O6.3e-8 & $5.0\times 10^5$ & $6.3\times 10^{-8} $ & 52.16 & 0.022 & --  & -- & 250  & --, Figure \ref{fig:verticalall}  \\
     S2e6O1.2e-7& $2.0\times 10^6$ & $1.2\times 10^{-7} $ & 48.01 & 0.27 & --& -- & 100  & Severe outflow  \\
     S2e6O1.2e-7\_z & -- &--& -- & -- & $8\times 8\times 4$ & $128\times 128\times 64$ & 400  & Turbulence, Figure \ref{fig:verticalall}  \\
     
     S2e6O2e-8\_res & -- &$2\times 10^{-8}$& 112.81 & 30.24 & $4\times 4\times 2$ & $256\times 256\times128$ &  300 & --, Figure \ref{fig:radiation_snapshot} - \ref{fig:chi_summary} \\

     \hline

     S2e4O6e-9$^*$ & $2.0\times 10^4$ & $6.0\times 10^{-9} $  &1.40 & $0.014$ & $8\times 8\times 2$ & $256\times 256\times 64$  & 20  & Fragmentation\\
     
     S2e4O2.5e-9 & -- & $2.5\times 10^{-9} $  &0.11 & $1.30$ & --  & -- & 3 & --\\
     
     S5e4O1e-8$^*$ & $5.0\times 10^4$ & $1.0\times 10^{-8} $ & 1.49 & 0.051 & -- & --& 15 & -- \\
      
     S5e4O8e-9 & -- & $8.0\times 10^{-9} $ & 0.56 & 0.20 & --  & --& 10 & --  \\
     
     S5e4O5e-9 & --&$5.0\times 10^{-9} $   &0.33 &1.28 & --&-- & 4 & --  \\
     
    S1e5O1.8e-8$^*$ & $1.0\times 10^5$&$1.8\times 10^{-8} $   &5.26 &0.031 & --&-- & 30 & --, Figure \ref{fig:S1e5O18e8snapshot}, \ref{fig:S1e5O18e8variable}  \\

    S1e5O1.5e-8 & -- &$1.5\times 10^{-8} $   &2.01 &0.11 & --&-- & 18 & --, Figure \ref{fig:S1e5O18e8snapshot}, \ref{fig:S1e5O18e8variable}  \\
    
    S1e5O1e-8 & -- &$1\times 10^{-8} $   &0.81 &0.73 & --&-- & 15 & --  \\
    
    S1e5O2e-9 & -- &$1.5\times 10^{-8} $   &2.82 &34.15 & --&--& 6 & --, Figure \ref{fig:S1e5O2e9snapshot} ,\ref{fig:S1e5O2e9variable}, \ref{fig:chi_summary}  \\
    
    S2e5O2.8e-8$^*$  & $2.0\times 10^5$ & $2.8\times 10^{-8} $ & 7.99 & 0.055 & -- & -- & 30&-- \\
     
    S2e5O1.5e-8  & -- & $1.5\times 10^{-8} $ & 1.86 & 1.07 & -- & -- & 10 &--\\
    
    S3e5O4e-9 & $3.0\times 10^5$ & $4\times 10^{-9} $ & 12.72 & 44.58 & -- & --  & 25 &--, Figure \ref{fig:S1e5O2e9snapshot} ,\ref{fig:S1e5O2e9variable}, \ref{fig:chi_summary}  \\
     
    S5e5O5.5e-8$^*$ & $5.0\times 10^5$ & $5.5\times 10^{-8} $ & 23.54 & 0.060 & -- & -- & 40 &-- \\
    
    S5e5O5e-8 & -- & $5.0\times 10^{-8} $ & 14.82 & 0.11 & -- & --  & 25 &-- \\
    
    S5e5O3e-8 & -- & $3.0\times 10^{-8}$ & 5.85 & 1.05 & -- & --  & 10 &-- \\

    S2e6O1e-7  & $2.0\times 10^6$ & $1.0\times 10^{-7} $  & 35.62 & 0.60 & -- & --  & 40 & -- \\ 
     
    S1e5O2e-9\_res & $1.0 \times 10^5$ & $2\times 10^{-9} $  &2.82 &34.15 & $4\times 4\times 2$ & $256\times 256\times128$ &  8 & --\\
    
    S3e5O4e-9\_res & --  &--& -- & -- & -- & -- &  30 & --\\

    \hline
    S1e5O2e-8  & $1.0\times 10^5$ & $2.0\times 10^{-8} $  & 10.23 & 0.014 & $8\times 8\times 2$ & $128\times 128\times 32$  &  200& Marginal \\
    
    S2e5O3.4e-8  & $2.0\times 10^5$ & $3.4\times 10^{-8} $  & 26.12 & 0.013 & -- & --  & 200 & -- \\
    
    S5e5O6e-8  & $5.0\times 10^5$ & $6.0\times 10^{-8} $  & 38.63 & 0.032 & -- & --  & 300& -- \\

     
     \hline
   \end{tabular}
   \caption{Summary of simulation input parameters. -- means ``as above.'' 
   Cases with $^*$ have corresponding runs with $2\times$ coarser resolution which demonstrate very similar properties; 
   we omit them in this table. }
   \label{tab:para}
  \end{table*}


\section{Results}
\label{sec:results}

\subsection{General Categorization}

We summarize initial parameters for all our runs in \autoref{tab:para}, 
and categorize outcomes into turbulent, 
fragmentation, and marginal cases. 
In addition to the control parameters 
$\Sigma, \Omega$, and the initial $\tau_{\rm cool},\prg$ 
(defined at initial $Q_T\approx 1.05$, see \S \ref{sec:initial}),
we also record the 
total length of simulations $t_{\rm final}$ in terms of $\Omega^{-1}$.

In the short-term {fragmentation} cases, 
compact density waves start to form within the first $\sim \tau_{\rm cool}$ and eventually break into bound clumps. 
Just a few $\Omega^{-1}$ after the initial formation of bound clumps, their central regions become extremely dense, and in this situation of spatially unresolved gravitational collapse the Riemann solver fails.
We practically terminate the simulation around this time and 
declare a runaway fragmentation. 

In other cases, 
growth of overdensity is limited by shear and structures are dispersed before they grow too compact.  In this situation, 
we only see transient clump formation in either 
a) the initial adjustment stage for the {gravito-turbulent} cases, 
or  
b) throughout the entire simulation for the {marginal} cases. 

In cases where there is not immediate gravitational runaway,
the disk usually settles into a quasi steady-state thermal equilibrium after the development of turbulence, which happens around a few cooling times. 
However, 
it can be difficult to determine a clean boundary between gravito-turbulence and fragmentation due to the existence of certain marginal cases. 
In these cases, bound clumps never exist for longer than a few orbital times, 
but because of large stochastic excursions the disk cannot maintain the steady excitation and disspation of turbulence needed in order for heating to balance cooling.   
The features of these scenarios will be described in detail below, 
but here we remark that without runaway collapse, both gravito-turbulent and marginal cases can both be run for much longer timescales, 
at least for a number of $\tau_{\rm cool}$. Nevertheless, 
all quasi-steady time-average values in \S \ref{sec:diagnostic} are only well-defined in the quasi-steady turbulence states.

\begin{figure*}[htbp]
\centering
\includegraphics[width=0.95\textwidth,clip=true]{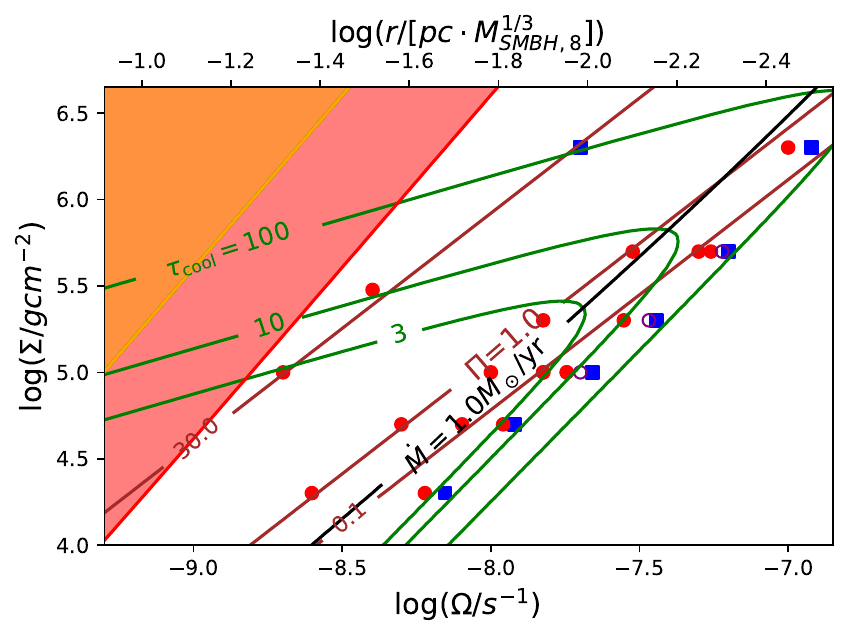}
\caption{Similar to \autoref{fig:contours_Pi_T}, \autoref{fig:contours_tcool}, \autoref{fig:contours_Mdot}, but each extra point represents outcome of simulations run with a set of $\Sigma, \Omega$ parameters.
Red dots and blue squares represent fragmentation and gravito-turbulent cases, 
respectively, 
while purple circles represented marginal cases. The definition of other contours and shaded regions are the same as in \autoref{fig:contours_Mdot}.}
\label{fig:tcool_data}
\end{figure*}

In the $(\Sigma, \Omega)$ plane, the outcome of our runs can be qualitatively summarized in \autoref{fig:tcool_data}, 
where each symbol represents $\geq 1$ simulation. 
Extra care should be taken for some points that are run with different resolution in the radiation-dominated regime, 
as elaborated in \S \ref{sec:rad_frag_results}. Generally, 
we observe that while fragmentation cases (red) and gravito-turbulent cases (blue) 
separated along the  $\tau_{\rm cool} \sim 3-5$ 
contour in the gas-pressure dominated limit to the lower right ($\Pi \lesssim 0.1$), 
the fragmentation boundary shifts to larger $\tau_{\rm cool}$ as one increases the radiation pressure fraction towards $\Pi=1.0$, with some marginal cases (purple) in between.  This indicates that it is more difficult to maintain gravito-turbulence against strong cooling in more radiation dominated disks. 
As we shall show below, 
this trend is more 
quantitatively defined by comparing the time averaged $\langle \tau_{\rm cool} \rangle_t$ and $\langle \prg'\rangle_t$ in the final states of gravito-turbulent cases. 

\subsection{The Fiducial Gravito-Turbulent Run}

We start by investigating classical gravito-turbulent states in the $\prg \ll 1$ limit. 
Such steady states are expected to self-regulate at $\langle \tau_{\rm cool}\rangle_t > 3 $ and $Q_T \gtrsim 1$ \citep[e.g.][]{Johnson2003,Rice2003}. To find a fiducial case in the classical gas pressure dominated limit, 
we start along the $\tau_{\rm cool} = 3$ contour at $\Sigma = 5\times 10^4 {\rm g/cm}^2$ on our parameter map (\autoref{fig:tcool_data}). 
Finding that $\Omega = 10^{-8}$s$^{-1}$ leads to fragmentation, 
we shift towards slightly larger $\Omega=1.2\times 10^{-8}$, with longer $\tau_{\rm cool}$.  This model, as indicated by the blue square in \autoref{fig:tcool_data}, evolves to reach a quasi-steady gravito-turbulent state.  
In principle, we could start from the far right where  $\tau_{\rm cool} \gg 1$ and then probe leftwards for a fragmentation boundary, 
but it would be expensive to run many gravito-turbulence simulations for a few times $\tau_{\rm cool}$. Instead, 
our strategy is to start from cases with the lowest $\langle\tau_{\rm cool}\rangle_t \sim \tau_{\rm cool}$ possible along each constant $\Sigma$ (or optical depth). 
This makes it possible to run simulations for timescales much longer than the local cooling time. 

\begin{figure*}[htbp]
\centering
\includegraphics[width=1\textwidth,clip=true]{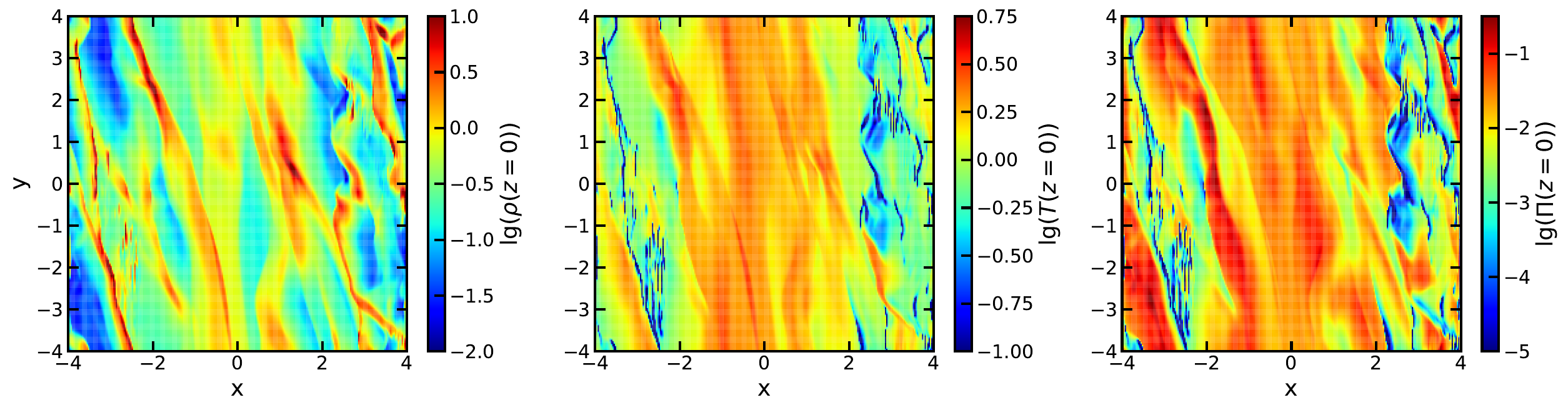}
\caption{Slices of midplane density, temperature, and radiation pressure fraction $\Pi$ in the fiducial simulation  at the final time ($t=150 \Omega^{-1}$). The density and temperature are normalized by initial midplane values for a $Q_T\sim 1.05$ disk. 
The length unit is $L_T = \pi^2 G \Sigma/\Omega^2$. }
\label{fig:fiducialsnapshot}
\end{figure*}

\begin{figure*}[htbp]
\centering
\includegraphics[width=0.5\textwidth,clip=true]{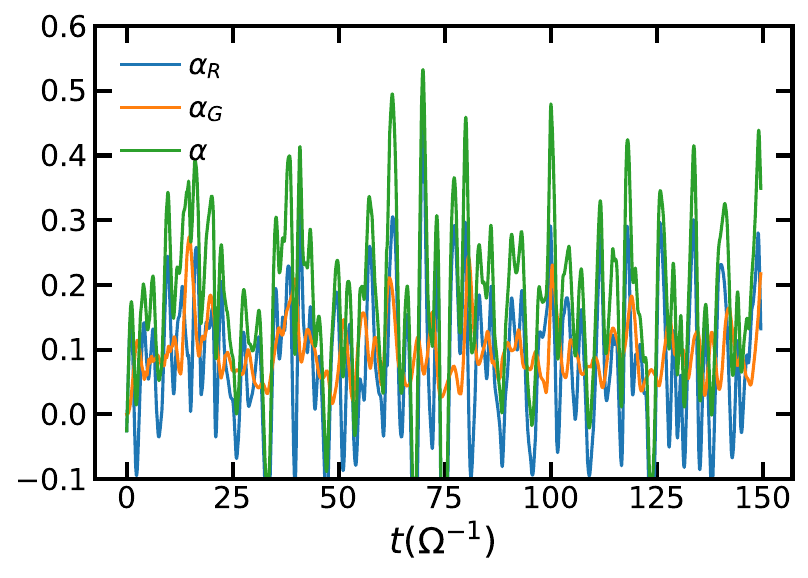}
\includegraphics[width=0.485\textwidth,clip=true]{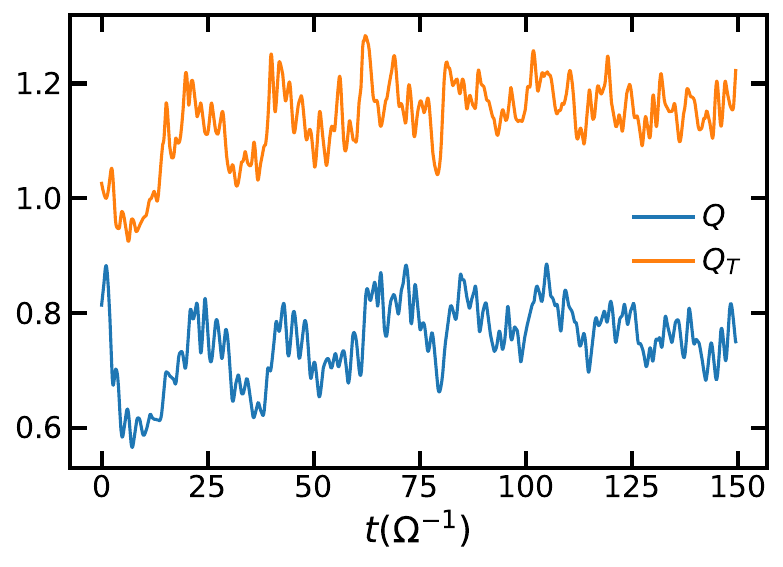}
\includegraphics[width=0.5\textwidth,clip=true]{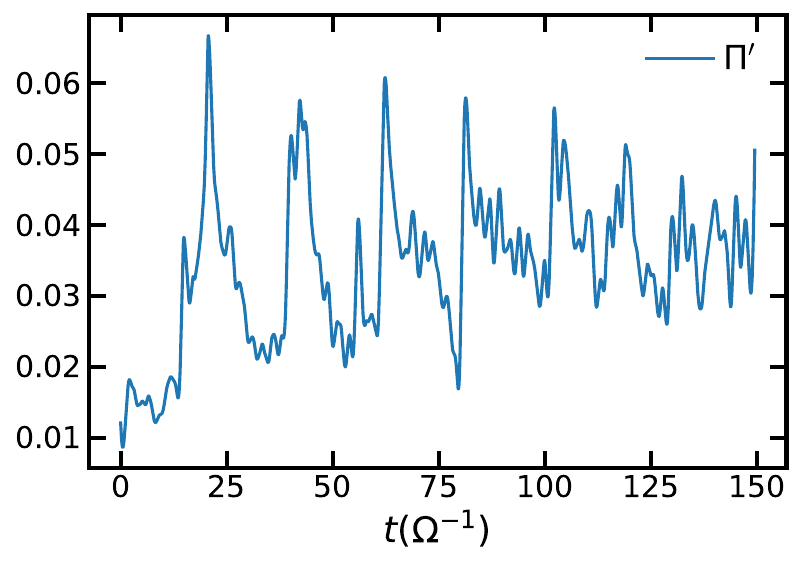}
\includegraphics[width=0.482\textwidth,clip=true]{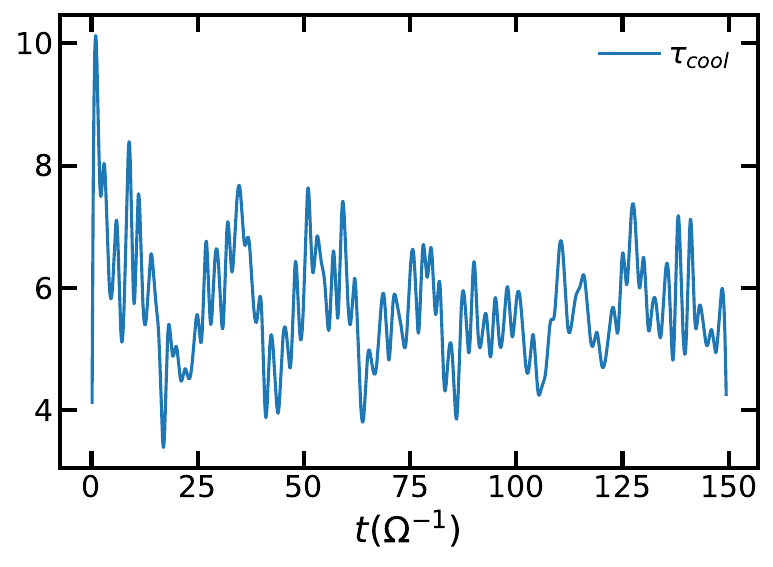}
\caption{The evolution of globally averaged variables $\alpha$, $Q_T$, $\Pi'$, and $\tau_{\rm cool}$ in our fiducial simulation. These average variables are defined in \S \ref{sec:diagnostic}.}
\label{fig:fiducialvariable}
\end{figure*}

\begin{figure}[htbp]
\centering
\includegraphics[width=0.45\textwidth,clip=true]{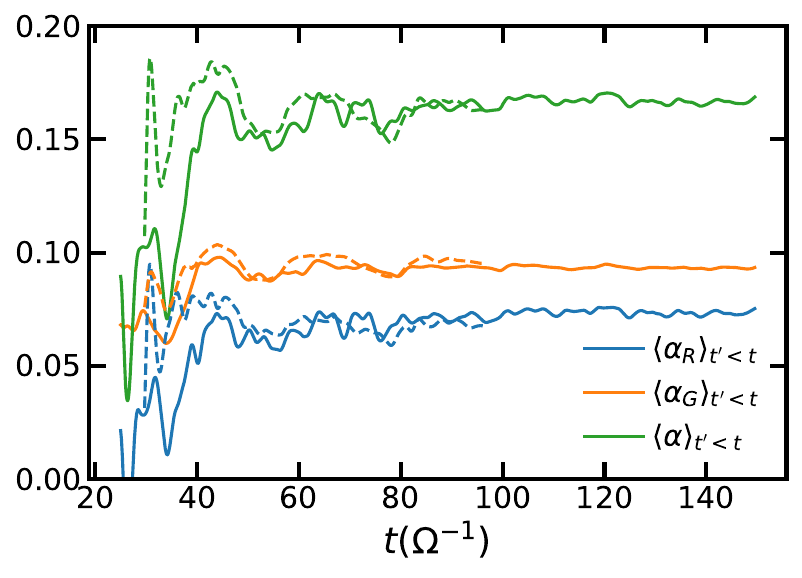}
\caption{Running time-average of stress parameters in the fiducial gravito-turbulent model (solid lines) and the doubled resolution run of the same model (dashed lines). 
}
\label{fig:alpharunningaverage}
\end{figure}

\begin{figure}[htbp]
\centering
\includegraphics[width=0.45\textwidth,clip=true]{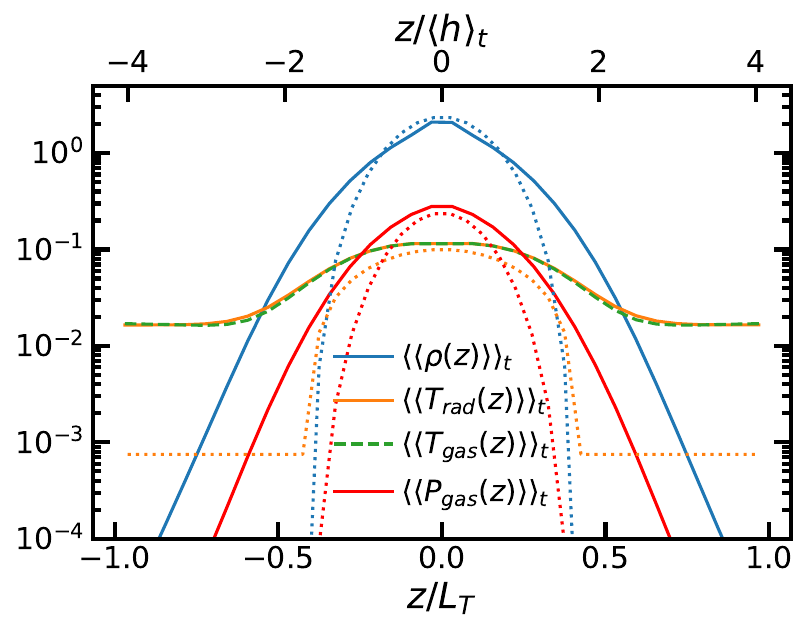}
\caption{Solid lines: the horizontally and temporally averaged vertical distribution of density, 
gas/radiation temperature, and gas pressure in code units for our fiducial run, after quasi-steady state is reached. 
Dotted lines show the initial vertical profiles of the run, corresponding to constant $\prg$. The top axis is normalized in terms of the averaged scale height.}
\label{fig:verticalfiducial}
\end{figure}

We present results of the fiducial case at $\Sigma = 5\times 10^4$g/cm$^2$, $\Omega = 1.2\times 10^{-8}$s$^{-1}$ in Figures \ref{fig:fiducialsnapshot}, \ref{fig:fiducialvariable}, \ref{fig:alpharunningaverage}, \ref{fig:verticalfiducial}. Snaphots of the midplane $\rho, T$ and $\Pi$ distribution at the end of the simulation $t=150 \Omega^{-1}$ are shown in \autoref{fig:fiducialsnapshot}, when gravito-turbulence has long since reached a quasi-steady state. 
Evolution of globally averaged variables $\alpha, Q_T, \prg', \tau_{\rm cool}$ {(definitions see \S \ref{sec:diagnostic})} collected at a cadence of $0.1\Omega^{-1}$ 
are shown in \autoref{fig:fiducialvariable}, 
from which we see that the disk undergoes a small initial cooling phase 
as $Q_T$ dips below 1 
before growth of stress parameters. 

The disk reaches a quasi-steady state after $t\sim 25 \Omega^{-1}$, 
and we run the simulation up to $t=150 \Omega^{-1}$.  This gives us 20-30 average cooling timescales $\langle \tau_{\rm cool}\rangle_t$ during our averaging interval $t_{\rm ave}=125\Omega^{-1}$, from 25-150 $\Omega^{-1}$, which is more than sufficient for running averages of fluctuating quantities to converge. 
For example, 
although it is easy to see that other global quantities have reached quasi-steady oscillation from \autoref{fig:fiducialvariable}, 
the stress parameter $\alpha$ and its components are fluctuating on a sub-orbital frequency with large amplitude. 
In \autoref{fig:alpharunningaverage} we show in solid lines 
the running averages of the stress parameters starting from $25\Omega^{-1}$ for the fiducial case (solid lines), 
which converge to steady values after a few tens of dynamical timescales. 
The heating produced when the total stress is $\langle \alpha \rangle_t \sim 0.17$ 
is able to balance the average cooling timescale of $\langle \tau_{\rm cool} \rangle_t \sim 5.62$, 
which is slightly longer than the initial value $\tau_{\rm cool} =4.56$. 
The final $\langle\Pi' \rangle $ reaches 0.034, 
which is $\sim$3 times larger 
than the initial value 0.013, 
suggesting a slightly hotter final state than the fiducial initial condition.

We also run the fiducial parameter model at higher resolution
and confirm that the averaged history outputs and average vertical profile converge with the standard resolution run. 
The running averages of stress parameters from the high resolution run \texttt{fiducial\_res} 
are shown in \autoref{fig:alpharunningaverage} as dashed lines 
(averaging starts at 30$\Omega^{-1}$ when disk enters a steady state), 
which converge to values similar to the standard run after 100$\Omega^{-1}$.
All other averaged vital variables are collected in \autoref{tab:para_turb} under the entry \texttt{fiducial} and \texttt{fiducial\_res}. 
The average $\langle \tau_{\rm cool}\rangle_t$ after 100 $\Omega^{-1}$ is slightly longer for the high-resolution run than the standard run, 
and the radiation fraction is slightly smaller.  
The computational cost for this high resolution run is about 3 Million CPU core hours. 
Given this high expense, for subsequent long-term simulations for gravito-turbulence states, 
we mainly adopt the standard resolution.

The time-averaged vertical distribution $\rho(z)$, 
radiation, gas temperature $T_{rad}(z), T_{gas}(z)$ 
and total gas pressure $P_{\rm gas}$ in our fiducial run are plotted in \autoref{fig:verticalfiducial}. 
The initial  $\rho$, $T_{gas} = T_{rad}$, and $P_{\rm gas}$ vertical profiles are plotted in dotted lines.
In steady-state, the gas and radiation are in thermal equilibrium, 
but the temperature and density profiles are generally more extended than the initial polytrope within the photosphere. 
The temperature and density gradient relaxes from our initial configuration and becomes much less steep around the photosphere. 
The temperature (radiation energy density) outside the photosphere region is isothermal, 
although it has little effect on the general gravito-turbulence since gas is tenuous in that optically-thin region. 
The average vertical $\langle \prg'(z)\rangle_t $ 
distribution within the photosphere is close to constant in our fiducial case (see \autoref{fig:verticalall}), 
not too far from our initial assumption.  However,
but we shall see this may not be valid for larger $\prg'$ runs 
when we compare a sequence of vertical profiles for gravito-turbulence cases in \autoref{fig:verticalall}, 
normalized by the midplane values of the averaged profiles themselves.

\subsection{Gas Pressure Dominated Gravito-Turbulent States}

Starting from our fiducial case which lies to the right of the stability boundary, we continue to explore the $\Sigma, \Omega$ plane as indicated with points shown in \autoref{fig:tcool_data}.
In the $\prg \ll 1$ limit, 
we identify another steady case at a smaller surface density $\Sigma=2\times 10^4$g/cm$^2$. 
The initial midplane temperature is 2000K so constant opacity is not in practice realistic, 
but this case does have $\langle \tau_{\rm cool}\rangle_t = 4.11$, 
which serves the purpose of validating consistency with the $\tau_{\rm cool}\gtrsim 3$ classical fragmentation criterion. 

Exploring along the other direction of expected $\tau_{\rm cool}$ contours with increasing $\prg$, 
we further identify 4 four gravito-turbulent cases, plotted as blue squares in \autoref{fig:tcool_data}. The parameters for all these runs are summarized in the first part of \autoref{tab:para}, 
and outcomes for averages of variables are summarized in \autoref{tab:para_turb}.
We perform time averages starting from $t=t_{\rm final}- t_{\rm avg}$  
to the end of the simulation $t_{\rm final}$, where $t_{\rm avg}$ is listed in \autoref{tab:para_turb}.

\begin{table*}
\centering
  \begin{tabular}{cc|cccccccccc}
     \hline\hline
     model & $ t_{\rm avg}\Omega$& $\langle Q_T \rangle_t$ & $\langle Q \rangle_t$ & $\langle \alpha\rangle_t$ & $\langle \alpha_G \rangle_t$ & $\langle \alpha_R \rangle_t$ & $\langle \tau_{\rm cool}\rangle_t$ & $\langle \Pi' \rangle_t$ & $\langle h\rangle_t$ & $\langle \tau_{\rm cool}\rangle_t/\tau_{\rm cool}(t=0)$ & $\langle \prg'\rangle_t/\prg(t=0)$\\

     \hline
     S2e4O7e-9 & 80 & 1.08 & 0.70 & 0.18 & 0.10 & 0.081 & 4.11 & 0.022 & 0.22 & 1.04 & 5.24\\
     
     fiducial & 125 &1.15 & 0.76 & 0.17 & 0.093 &0.075 & 5.62 & 0.034 & 0.24 & 1.23 & 2.62\\
     
     fiducial\_hi & 70 & 1.11 & 0.73&0.16 &0.095 & 0.069 & 6.48 & 0.029 & 0.23 & 1.42 & 2.23\\
     
     S1e5O2.2e-8& 80 & 1.22 & 0.79 & 0.17 & 0.087 & 0.086 & 6.44 & 0.035& 0.25 & 0.33 & 5.15 \\
     
     S2e5O3.6e-8 & 150 & 1.25 & 0.80 &0.10 & 0.047& 0.053 & 12.08 & 0.048 & 0.25 & 0.32 & 5.71\\
     
     S5e5O6.3e-8 & 150 & 1.36 & 0.87 & 0.048 & 0.025 &0.033 & 20.52 & 0.12 & 0.27 & 0.39 & 5.45\\
     
     S2e6O1.2e-7\_z & 150 & 1.42 & 0.94 & 0.023& 0.012& 0.011& 42.17 & 1.04 &0.30 & 0.88 & 3.85\\
     \hline
     S2e6O2e-8\_res & 300 & 1.19 & 0.84 & 0.0062 & 0.0018 &0.0044 & 150.63 & 38.31 & 0.23 & 1.34 & 1.27\\
     \hline
   \end{tabular}
   \caption{Summary of outcomes in gravito-turbulent cases. Initial conditions are given in the upper portion of \autoref{tab:para}. }
   \label{tab:para_turb}
  \end{table*}

\begin{figure*}[htbp]
\centering
\includegraphics[width=0.485\textwidth,clip=true]{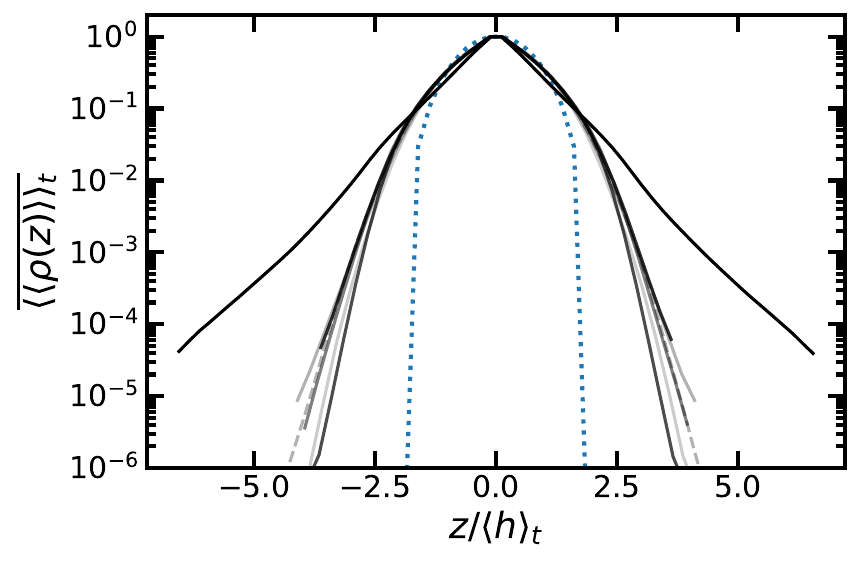}
\includegraphics[width=0.485\textwidth,clip=true]{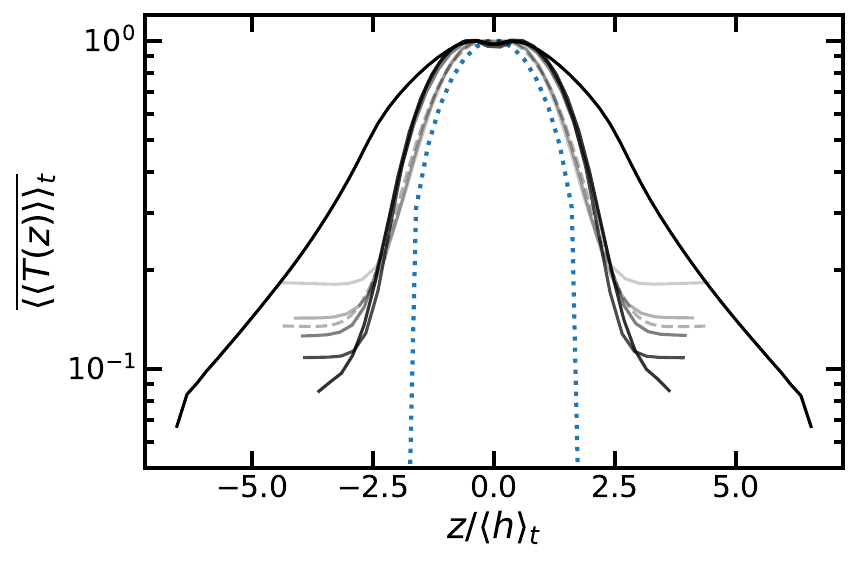}
\includegraphics[width=0.485\textwidth,clip=true]{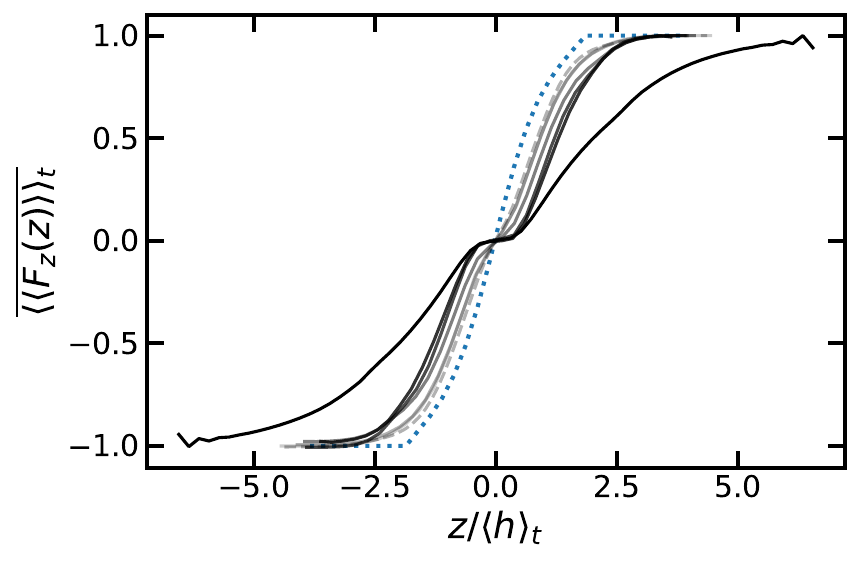}
\includegraphics[width=0.485\textwidth,clip=true]{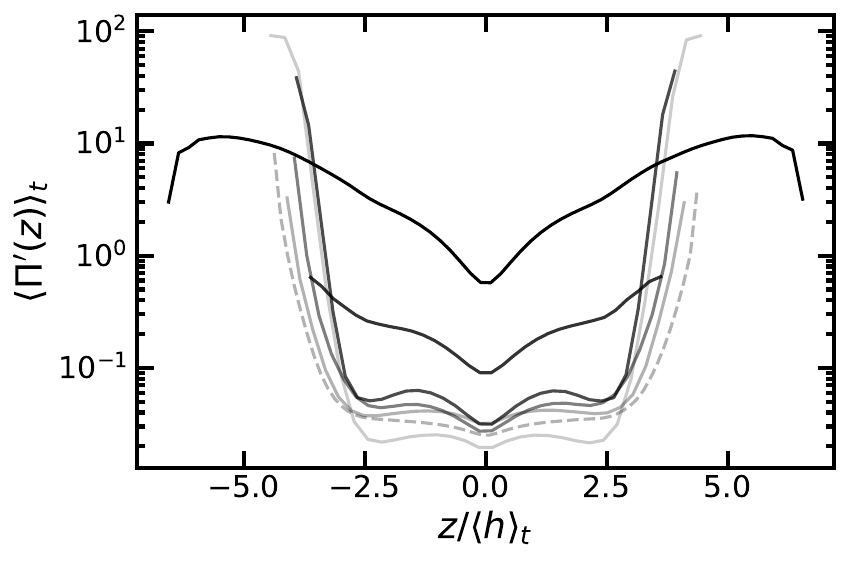}
\includegraphics[width=0.487\textwidth,clip=true]{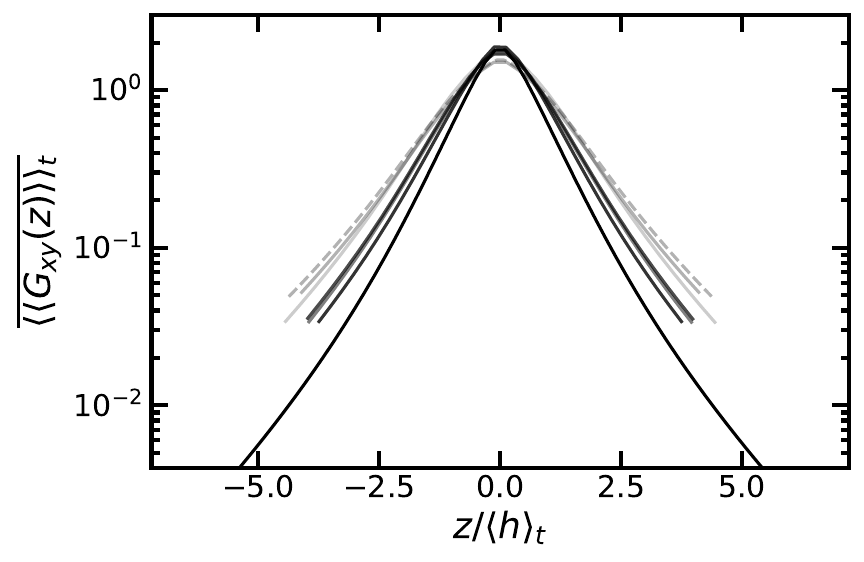}
\includegraphics[width=0.487\textwidth,clip=true]{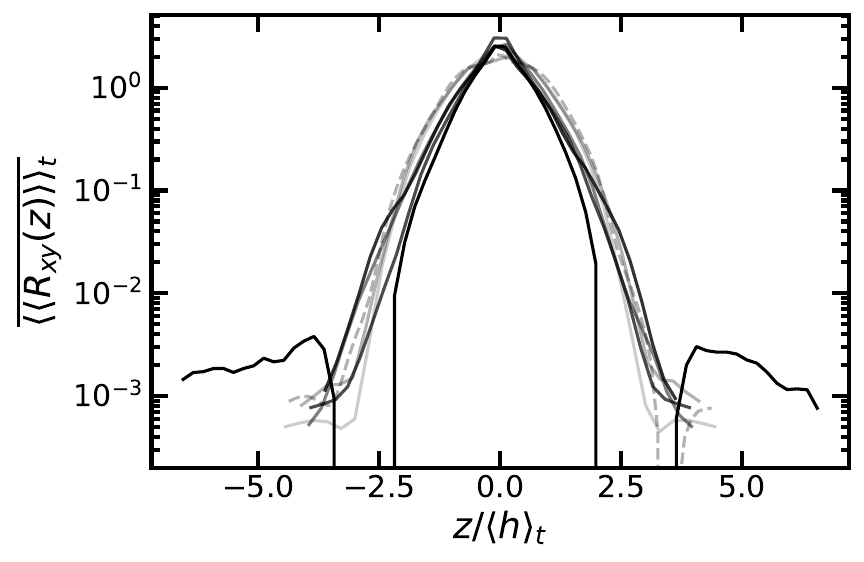}
\caption{Horizontally and temporally averaged vertical distributions of density, temperature, vertical radiative flux, pressure ratios, and stress functions in the quasi-steady gravito-turbulent cases. Darker curves represent cases with larger $\Sigma$ and radiation pressure fraction. Dotted lines show the adopted initial vertical distribution. All profiles are normalized (as indicated by overline); see text. }
\label{fig:verticalall}
\end{figure*}

In \autoref{fig:verticalall} we show the averaged vertical distributions of density, temperature, 
vertical radiation flux as well as 
pressure ratios of these cases, 
with the vertical axis normalized in terms of average scale height $\langle h\rangle_t$. 
Density and temperature profiles $\overline{\rho}, \overline{T}$ 
are normalized by the midplane value, 
and $\overline{F_{z}}$ is normalized by $|F_{z,max}|$ at the disk surface. 
Darker curves represent cases with larger time-averaged $\prg'$, and in this case, 
also larger optical depth $\kappa \Sigma/2$. 
The fiducial high resolution case is plotted with dashed lines. 
All the solid lines represent cases run with standard resolution. 

Most cases adopt a standard box size, 
with the exception of \texttt{S2e6O1.2e-7\_z} which applied $8 L_T\times 8L_T \times 4 L_T$ with doubled grid number in the vertical axis 
(ergo same resolution). 
The standard box size cases all have $\langle \prg' \rangle_t < 1$, 
and the vertical distribution of quantities share similar profiles plotted in terms of the scale height. 
Within $\pm 2\langle h \rangle_t $, 
these profiles are similar to the initial polytrope, 
with $\langle \Pi' \rangle_t$ undergoing little change within one scale height, although the normalization $\langle h\rangle_t$ 
(measured in units $L_T = \pi l_*$) 
itself is an increasing function of $\Sigma$, 
consistent with an increasing $\langle Q_T\rangle_t$ in quasi-steady state. 
The radiation pressure becomes dominant outside the photosphere, 
where temperature and gas density are low and cannot affect disk dynamics.

Run \texttt{S2e6O1.2e-7\_z} with radiation comparable to gas pressure stands out 
because its profile becomes much more extended than the initial condition, 
and we have to apply a larger box to accommodate the quasi-steady profile. 
In fact, 
the box size was adjusted after we found in run \texttt{S2e6O1.2e-7} 
that the standard box size leads to severe mass outflow.  
This is why in \autoref{fig:verticalall} for all other cases only $L_T\langle h\rangle_t^{-1}\lesssim 5 $ scale heights are covered on either side 
but for \texttt{S2e6O1.2e-7\_z} we cover $2L_T \langle h\rangle_t^{-1}\sim 7$ scale heights on each side. 
Compared with other runs at smaller $\langle \prg' \rangle_t$, the vertical temperature profile does not drop as rapidlhy as a function of $z in$ \texttt{S2e6O1.2e-7\_z}.
Towards the boundaries, 
the density profile is less steep, 
and since temperature also drops off more slowly, 
$\prg'(z)$ has a large deviation from the midplane value. 

The stress parameters' vertical profiles are plotted in the bottom panels of Figure \ref{fig:verticalall}.
For this purpose, they are normalized by $\alpha  \langle \langle P_{\rm gas} + P_{\rm rad} \rangle \rangle_t(z=0)$ (as indicated by the overline) such that $
\int \overline{
\langle\langle R_{xy}(z)\rangle \rangle_t}
+\overline{
\langle
\langle G_{xy}(z)\rangle \rangle_t} 
{\rm d}z = 1$. 
We see that the extended temperature profile of \texttt{S2e6O1.2e-7\_z} is consistent with having velocity fluctuations 
which give rise to non-negligible Reynolds stress outside the photosphere, 
while in other cases both gravitational and Reynolds stress are appreciable only near the midplane.

\begin{figure}[htbp]
\centering
\includegraphics[width=0.47\textwidth,clip=true]{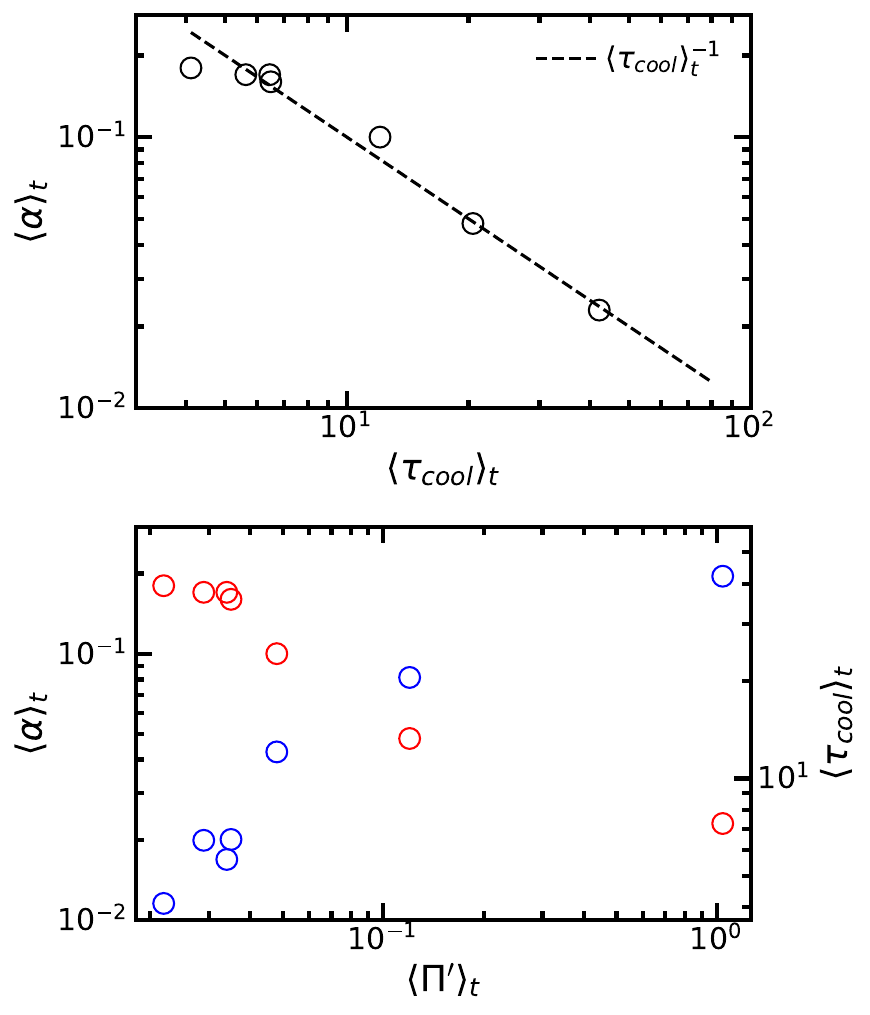}
\caption{Averaged turbulent parameters, cooling times and pressure ratios in the gravito-turbulent states. Upper panel: $\langle \alpha \rangle_t$ plotted against $\langle \tau_{\rm cool} \rangle_t$ to demonstrate heating-cooling equilibrium; Lower panel: $\langle \alpha \rangle_t$ ({\it red circles}) and $\langle \tau_{\rm cool} \rangle_t$ ({\it blue}) plotted against $\langle \Pi'\rangle_t  = \langle\langle P_{\rm gas}\rangle\rangle_t/\langle\langle P_{\rm rad}\rangle\rangle_t $ to show the effect of radiation pressure on the stability boundary.}
\label{fig:boundaryalpha_Prg}
\end{figure}

Before discussing outcomes of fragmentation, 
we can already gain some understanding of the effect of $\prg'$ from these quasi-steady states along the stability boundary. 
We plot the time-averaged $\langle \alpha \rangle_t$ as functions of $\langle  \tau_{\rm cool}\rangle_t$ and $\langle \prg'\rangle_t$ in \autoref{fig:boundaryalpha_Prg}. 
The upper panel demonstrates energy balance since $\alpha \approx \tau_{\rm cool}^{-1}$. 
Furthermore, 
we expect these variables to roughly trace out the shortest cooling timescale the disk could maintain stability against, 
or the largest turbulence strength the disk could possibly support. 

From the lower panel of \autoref{fig:boundaryalpha_Prg}, 
we see that for $\prg' \ll 1$ we recover the classical boundary where $\alpha$ has a limit of $\lesssim 0.2$,  
while $\tau_{\rm cool} \gtrsim 5$. 
As we increase $\prg'$ at higher $\Sigma$, 
we expect the effect of radiation pressure to extend the instability threshold to larger cooling time and smaller turbulence. 
Previous results from \citet{Jiang11} suggest that when $\prg' > 1$ the maximum quasi-steady stress $\alpha$ a disk could generate starts to deviate from $\sim 0.3$, decreasing down to $\lesssim 0.01$ at $\prg' \gtrsim  30$. 
In our simulations, 
we observe that $\prg' \sim 0.1$ is already sufficient to modify the stability boundary significantly such that the disk can only support $\alpha\lesssim 0.05$ with $\tau_{\rm cool} \gtrsim 20$. 
When $\prg'$ reaches order unity, the gravitationally-driven stress 
has dropped to $\alpha\sim$ 0.02, 
comparable to the level that can be provided by MRI \citep{Beckwith2011,Simon2012}. Although in this preliminary study, we have too few simulation data points to produce a robust empirical prescription for this dependence, 
simply extrapolating from the current trend towards $\langle \prg' \rangle_t$ of order 10, 
we expect a maximum gravitationally-driven turbulent stress $\alpha < 0.01$ for the highly radiation dominated regime. 
As discussed in \autoref{sec:formulation}, such a configuration may not be realistic since MRI can already provide sufficient heating to turn off GI
and the problem reduces to a thin $\alpha$ disk model. 
Further implications are discussed in \S \ref{sec:discussion}.

\subsection{Gas Pressure Dominated Fragmentation Cases}
\label{sec:gas_frag_results}

\begin{figure*}[htbp]
\centering
\includegraphics[width=0.9\textwidth,clip=true]{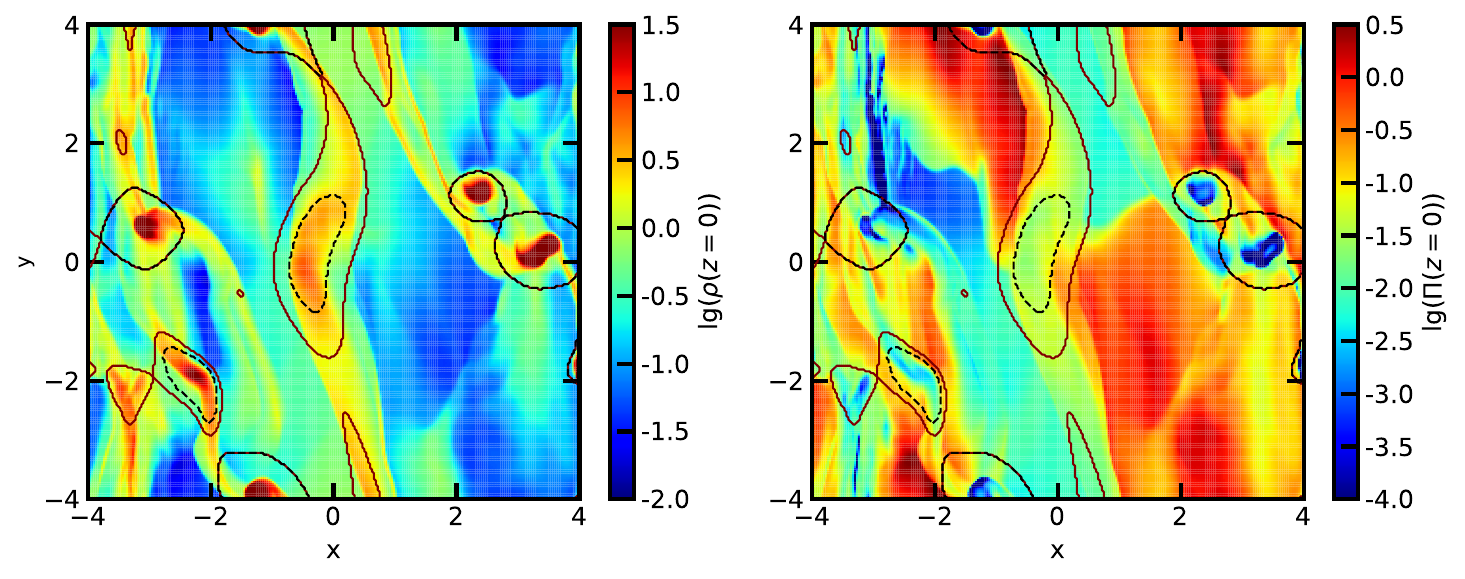}
\includegraphics[width=0.9\textwidth,clip=true]{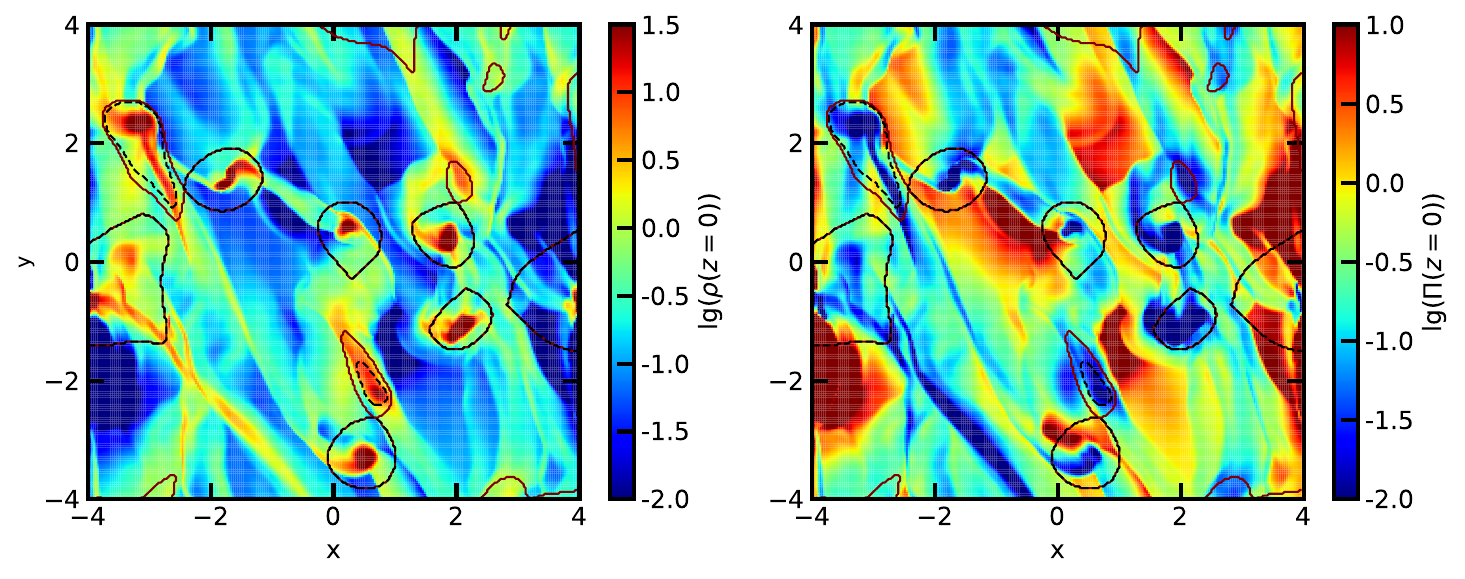}
\caption{Snapshots of midplane density and $\prg$ for two exemplary gas-dominated fragmentation cases with final $\prg' \lesssim 0.1$. Top panels show \texttt{S1e5O1.8e-8}, bottom panel \texttt{S1e5O1.5e-8}.}
\label{fig:S1e5O18e8snapshot}
\end{figure*}

Simulation cases that lead to fragmentation are plotted as red dots in \autoref{fig:tcool_data}. 
After roughly an initial $\tau_{\rm cool}$, 
radiative cooling leads to development of bound objects. 
We plot in \autoref{fig:S1e5O18e8snapshot} the midplane density and $\prg$ distribution for exemplary fragmentation cases \texttt{S1e5O1.5e-8} and \texttt{S1e5O1.8e-8} after the formation of bound objects. 
The brown contours indicate boundaries for 
HBPs,
while dashed black contours indicate bound regions (see \S \ref{sec:bound}). 
The overlapping of these contours means entire HBPs 
are bound, satisfying the HBR condition.

\begin{figure}[htbp]
\centering
\includegraphics[width=0.44\textwidth,clip=true]{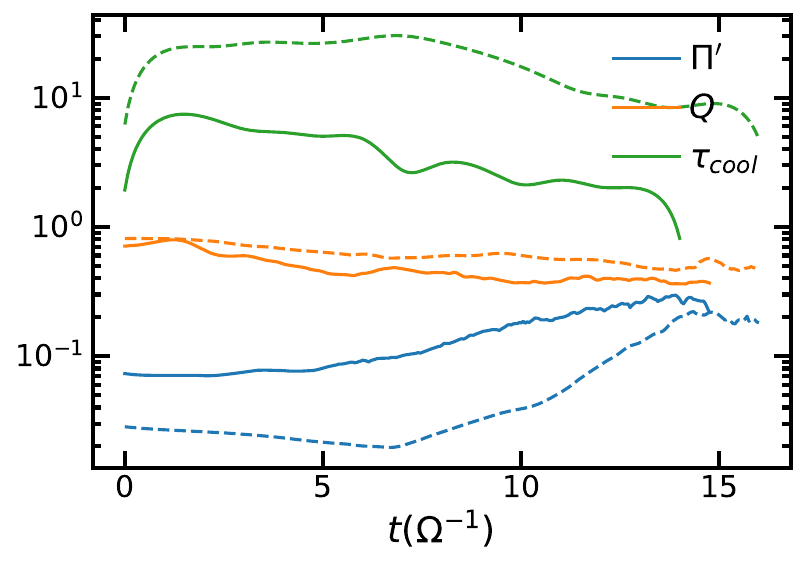}
\caption{Evolution of average variables $\prg'$, $Q$, and $\tau_{\rm cool}$ in cases \texttt{S1e5O1.8e-8} (dashed lines) and \texttt{S1e5O1.5e-8} (solid lines).}
\label{fig:S1e5O18e8variable}
\end{figure}

The evolution of spatially averaged variables $\prg', Q, \tau_{\rm cool}$ 
for these cases are plotted in \autoref{fig:S1e5O18e8variable}. 
both the initial $\prg$ 
and the final average $\prg'$ at the point of fragmentation is $\sim 0.1$; 
therefore the overall conditions are of gas pressure-dominated fragmentation. 
However, 
the distribution of $\rho$ and $\prg$ become quite inhomogeneous in the midplane, with low density regions where radiation pressure dominates appearing, 
while the bound regions cool off to become local minima of the radiation pressure fraction. 
Over the first cooling timescale, 
$Q$ nearly monotonically decreases, but even as the disk becomes gravitationally unstable it cannot generate enough turbulence to balance cooling. 
Once bound fragments form, 
no quasi-steady state can be achieved.
The overall behavior of $\tau_{\rm cool}$ is also similar in the two cases, 
with an initial increase in a brief phase of steady cooling, 
followed by a secular decrease after turbulence develops.
While the case \texttt{S1e5O1.5e-8} has a $ \tau_{\rm cool} <1$ upon the point of fragmentation,
the case \texttt{S1e5O1.8e-8} fragmented at $\tau_{\rm cool} \sim 7$;  
this implies that $\prg' \sim 0.1$ can already prevent the disk from maintaining a turbulence of $\alpha \sim 0.1$.
Other fragmentation cases with large surface density on the $\prg'<1$ side show similar behavior, 
with larger $\prg'$ runs being able to fragment at larger $\tau_{\rm cool}$. Specially, 
the run \texttt{S2e61e-7} (top-most red dot in \autoref{fig:tcool_data}) undergoes collapse at $\tau_{\rm cool} \sim 30$.

Despite the increase of critical cooling time at larger $\prg'$ as consistent with \autoref{fig:boundaryalpha_Prg}, 
the outcome in fragmentation cases with non-negligible $\prg'<1$ are still similar to those well-studied in the gas-pressure dominated Jeans instability theory.
Formation of bound regions has characteristic length scale $\sim L_T$ \citep[][see their \S 3.2]{GoodmanTan2004}, with \textit{initial} masses comparable to the Jeans mass $\Sigma L_T^2$. 
However, 
we do not follow long term development of these clumps in this paper.  We 
therefore cannot determine if merger and accretion 
renders larger final masses when structures become dense enough for nuclear fusion or core-collapse, or if alternatively the initial collapsing objects fragment into lower masses.  
While our current simulations do not permit us to reach any conclusions regarding
initial mass function for star formation in AGN disks,  \S \ref{sec:discussion} provides some rough estimates of initial fragment masses.

We have remarked that there are also marginally non-fragmentating simulations along the stability boundary that never settle into a steady-state.
In these runs, 
each time the disk 
cools down to a relatively small $Q \sim 0.5$ and 1-2 dense clumps begin to form in the midplane, 
strong in-homogeneous turbulence develops abruptly and heats the disk up to $Q \gtrsim 1$ again  within just a few dynamical timescales. 
As the disk reaches a hotter state, 
the outburst of turbulent heating dies out along with dispersal the bound clumps, but subsequently  
the disk cools down again with a small background turbulence unable to fully balance radiative cooling. 
We typically see $2-3$ of these cycles in our marginal cases before we stop the simulation, and without running for up to a thousand dynamical timescales there is no way to know whether the disk can reach a quasi-steady turbulent state. 
Since these models have quite distinctive behavior, however, 
we identify them as a third ``marginally unstable'' scenario (purple open circles in \autoref{fig:tcool_data}), but do not discuss them in detail. 
If these cases can stabilize given sufficient simulation time at larger $\prg'$ than our gravito-turbulent cases, 
then \autoref{fig:boundaryalpha_Prg} should be seen as a conservative estimate of the critical $\alpha$, 
suggesting that there is still room too support slightly larger quasi-steady turbulence against cooling.

\subsection{Radiation Pressure Dominated Fragmentation Cases}
\label{sec:rad_frag_results}

Moving further left in the $(\Sigma, \Omega)$ parameter space (see \autoref{fig:tcool_data}), 
there is  a group of runs along the initial $\prg =1$ contour that fragmented wat $\prg' \gtrsim 1$ and behave similarly ato those described in \ref{sec:gas_frag_results}. 
To better compare with the gas pressure dominated fragmentation outcomes, 
we present result from fragmenting runs  \texttt{S3e5O4e-9} and \texttt{S1e5O2e-9} with final $\prg' \sim 30$. 
The former case has $\tau_{\rm cool} \sim 10$, but boundary values from \autoref{fig:boundaryalpha_Prg} suggests that in the case of $\prg' \sim 0.1$ we already require $\tau_{\rm cool}> 10$ to stabilize, 
so for much larger $\prg'$, 
fragmentation is expected. 
The latter case has a cooling time within the $\tau_{\rm cool} \sim 3$ boundary so it would fragment even by classical standards, 
and we expect it to be even more susceptible to fragmentation in the radiation dominated situation. 

\begin{figure*}[htbp]
\centering
\includegraphics[width=0.9\textwidth,clip=true]{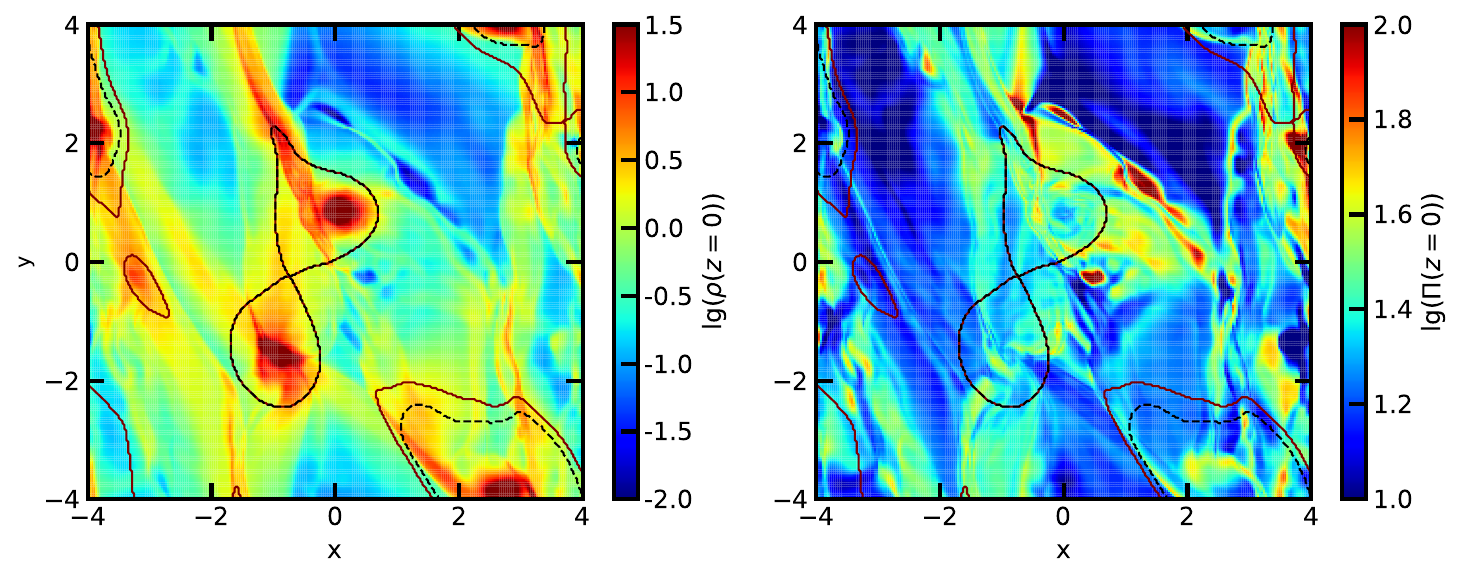}
\includegraphics[width=0.9\textwidth,clip=true]{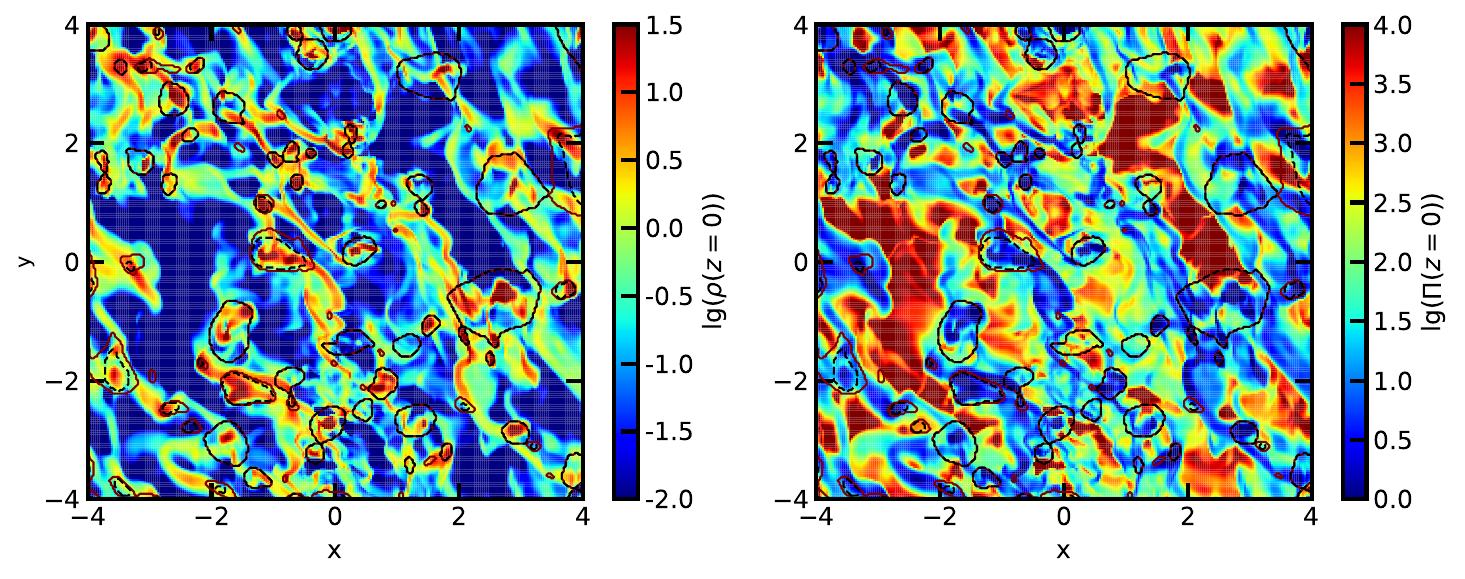}
\caption{Midplane slices of $\rho$ and $\prg$ for radiation-dominated fragmentation cases \texttt{S3e5O4e-9} (top) and \texttt{S1e5O2e-9} (bottom).}
\label{fig:S1e5O2e9snapshot}
\end{figure*}

\begin{figure}[htbp]
\centering
\includegraphics[width=0.44\textwidth,clip=true]{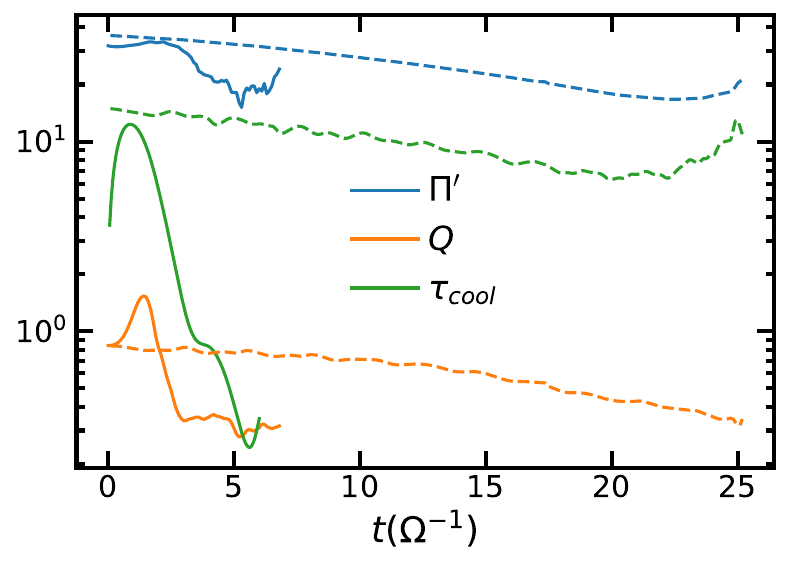}
\caption{Evolution of average variables $\prg'$, $Q$, and $\tau_{\rm cool}$ in cases \texttt{S3e5O4e-9} (dashed lines) and \texttt{S1e5O2e-9} (solid lines). 
}
\label{fig:S1e5O2e9variable}
\end{figure}

From \autoref{fig:S1e5O2e9snapshot} and \autoref{fig:S1e5O2e9variable} 
(plotted in the same manner as \autoref{fig:S1e5O18e8snapshot} and \autoref{fig:S1e5O18e8variable}), 
we observe that the decrease in $Q$ also become irreversible up to the point of fragmentation. 
The result from \texttt{S3e5O4e-9} 
reconfirms 
that in the radiation dominated regime, 
the classical criterion no longer applies since $\tau_{\rm cool} \gtrsim 10 $ still cannot prevent fragmentation.
The initial fragment sizes are still of order $L_T$, 
meaning that the pressure within these fragments is supported by both gas and radiation.

For run \texttt{S1e5O2e-9}, 
a final $\tau_{\rm cool} < 1$ is accompanied by rapid fragmentation after $\sim 5 \Omega^{-1}$.  
A notable feature is that the width of density waves upon fragmentation, 
and the size of subsequent clumps 
(as seen in \autoref{fig:S1e5O2e9snapshot}) are distinctively smaller than in other fragmentation cases. 

In the midplane where fragments first appear, 
the Jeans length can be expressed as
\begin{equation}
    L_J = c_{s} \left(\dfrac{\pi }{G\rho_0}\right)^{1/2} = \dfrac{c_s \Omega}{\pi G \Sigma} \left(\dfrac{\Omega^2}{\rho_0 \pi G }\right)^{1/2} L_T,
\end{equation}
where the characteristic sound speed $c_s \approx Q_T \pi G \Sigma/\Omega $ and midplane density $\rho_0 \approx   \Omega^2/(2 \pi G Q)$ will result in $L_J \approx Q_T \sqrt{2Q} L_T$. 
While we consistently observe bound regions with typical size/spacing of $L_J \lesssim L_T$ in the afore-mentioned fragmenting runs, 
for \texttt{S1e5O2e-9} in the $\tau_{\rm cool} < 1$ limit the size instead 
becomes more comparable to the gas-pressure Jeans length:
\begin{equation}
    L_{J,gas} = c_{s,gas} \left(\dfrac{\pi }{G\rho_0}\right)^{1/2} \approx \dfrac{L_J}{\left(1+\Pi'\right)^{1/2}}<  L_{J}.
\end{equation}
Here, $c_{s,gas}$ is the sound speed from gas pressure only. 
The $\prg$ distribution also suggests for this particular case, 
radiation quickly escapes bound clumps and they are supported 
exclusively by gas pressure. 
In \texttt{S3e5O4e-9}, 
there are also small filaments of density fluctuations on the scale of $L_{J,gas}$, but it is still the total pressure that determines the characteristic fragment size. 
We performed two additional runs with higher resolution \texttt{S3e5O4e-9\_res} and \texttt{S1e5O2e-9\_res} 
(smaller horizontal box size to save computation time) 
in which the respective $L_{J,gas}$ at $Q\sim 1$ is properly resolved by $\sim 10$ grid zones. 
Since we never reach a steady state 
and upon fragmentation $Q$ may reach as low as $\sim 0.3$, 
such resolution may be necessary. 
Nevertheless, 
in the high resolution runs
the qualitative fragmentation outcomes have not changed and the characteristic fragment size difference is still apparent. 
This dichotomy may be explained by a radiation diffusion rate criterion, 
separate from the fragmentation/stability boundary,
which we will elaborate in \S \ref{sec:diffusion}. 

\subsection{Radiation Pressure Dominated Gravito-turbulent Case}
\label{sec:radiation_turbulence}

\begin{figure*}[htbp]
\centering
\includegraphics[width=1\textwidth,clip=true]{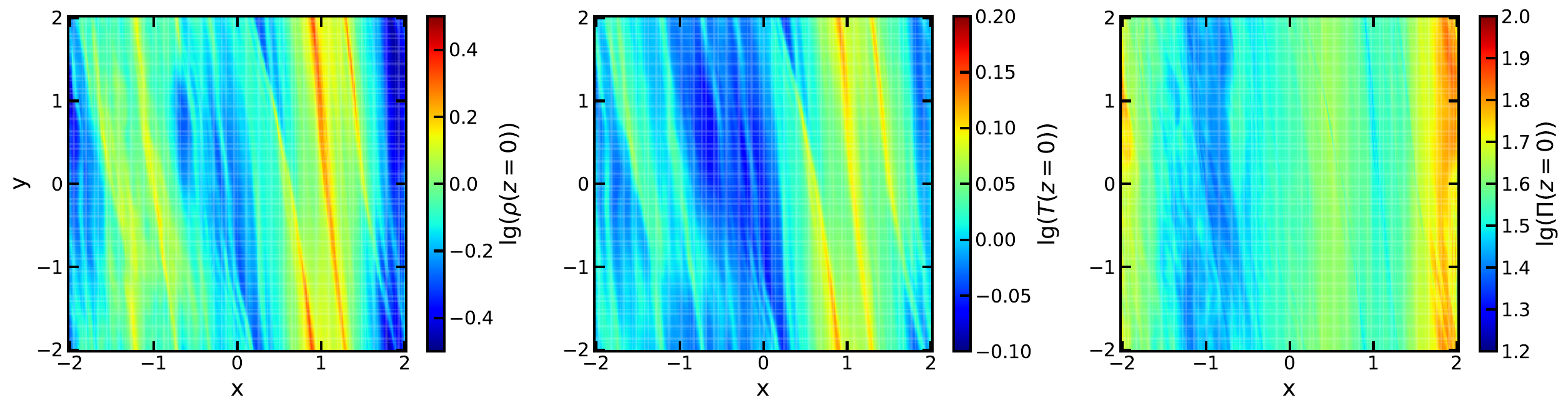}
\caption{Slices of midplane density, temperature and radiation pressure fraction $\Pi$ distribution 
at the final time ($t=400 \Omega^{-1}$) in radiation-dominated gravito-turbulent model \texttt{G2e6O2e-8\_res}. The density and temperature are normalized by initial values. The length unit is $L_T = \pi^2 G \Sigma/\Omega^2$. }
\label{fig:radiation_snapshot}
\end{figure*}

\begin{figure}[htbp]
\centering
\includegraphics[width=0.45\textwidth,clip=true]{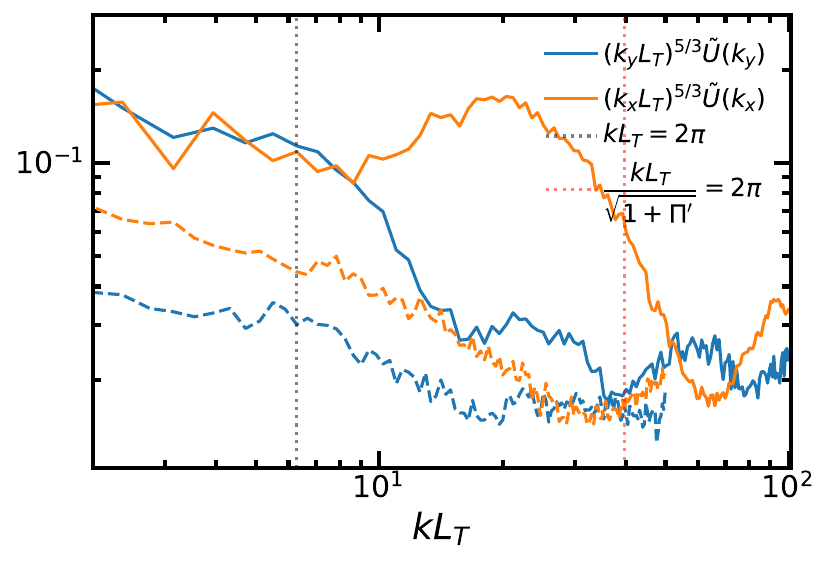}
\caption{Power spectrum of midplane total internal energy $U$ for gas pressure dominated gravito-turbulent case \texttt{fiducial} (dashed lines), versus radiation pressure dominated gravito-turbulent case \texttt{G2e6O2e-8\_res} (solid lines), scaled with $k^{5/3}$. Blue lines and orange lines indicate spectrum in $x$ and $y$ directions, respectively. $\tilde{U}(k) ={U}(k)/\sum_k U({k}) $ is the normalized spectrum density. }
\label{fig:spectrum_comparison}
\end{figure}

The one exceptional radiation dominated case in \autoref{fig:tcool_data} is \texttt{S2e6O2e-8\_res},
which has an initial dimensionless cooling time $\tau_{\rm cool} \sim 100$. 
For this model with extremely weak radiative cooling, 
the simulation reaches a 
quasi-steady state with a large radiation pressure fraction, analogous to the previous gas pressure dominated cases \texttt{S1e5O2e-9} and \texttt{S3e5O4e-9}.
Because it is safest
to resolve the gas Jeans length for such simulations, 
we choose a small box size of $4\times 4\times 2 L_T^3$ with high resolution $256\times 256\times 128$, 
similar to \texttt{S1e5O2e-9\_res} and \texttt{S3e5O4e-9\_res} which converge with their low resolution counterparts in their fragmentation outcomes. 
We run this simulation for 400 orbits without runaway gravitational collapse. 
The computational cost for this run is about 10 Million CPU core hours. 

Snapshots of the midplane $\rho, T, \Pi$ distribution 
are shown in \autoref{fig:radiation_snapshot}. 
Note that the simulation domain is smaller than in \autoref{fig:fiducialsnapshot}. Compared to the fiducial gas pressure dominated simulation, 
the turbulence level for \texttt{S2e6O2e-8\_res} is very low 
and density/temperature fluctuations are considerably smaller. 
In addition to structures on the Jeans scale $\sim L_J$, there are also much 
smaller scale fluctuations in the $x$ direction, 
possibly related to $L_{J, \rm gas}$ as previously discussed . 

To quantify the structure in \texttt{S2e6O2e-8\_res}, 
we plot the midplane energy spectrum \footnote{We employ  the total internal energy $U$ for Fourier analysis because the time fluctuation in its spectrum is smaller compared to other variables.} 
in \autoref{fig:spectrum_comparison} with solid lines. 
Additionally, we plot the spectrum from the fiducial simulation \texttt{fiducial} with dashed lines. 
The spectra are multiplied by a factor of $k^{5/3}$ to indicate the break wavenumbers clearly \citep[e.g.][Figure 5]{Booth2019}. 
We observe that in gas pressure dominated model \texttt{fiducial} there are spectral breaks in both $x$ and $y$ directions around $k L_T\sim 2\pi$, 
in the radiation dominated run the $U(k_x)$ spectrum does extend to $\sim k L_T/\sqrt{1+\langle \Pi'\rangle_t}\sim 2\pi$ before breaking.
We briefly discuss the origin of these modes in \S \ref{sec:diffusion}.

\begin{figure}[htbp]
\centering
\includegraphics[width=0.45\textwidth,clip=true]{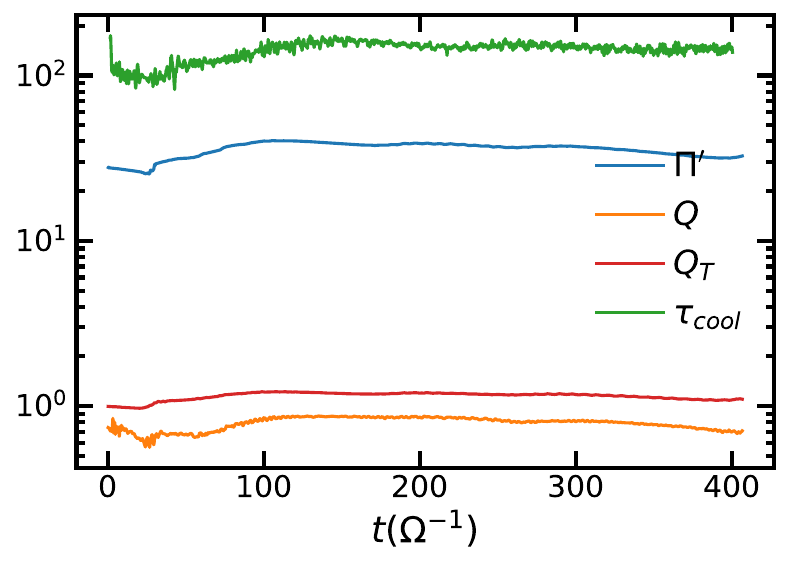}
\caption{The evolution of globally averaged variables $Q$, $Q_T$, $\tau_{\rm cool} $, and $\Pi'$ in the radiation dominated gravito-turbulent simulation \texttt{S2e6O2e-8\_res}. }
\label{fig:marginalvariable}
\end{figure}

\begin{figure}[htbp]
\centering
\includegraphics[width=0.44\textwidth,clip=true]{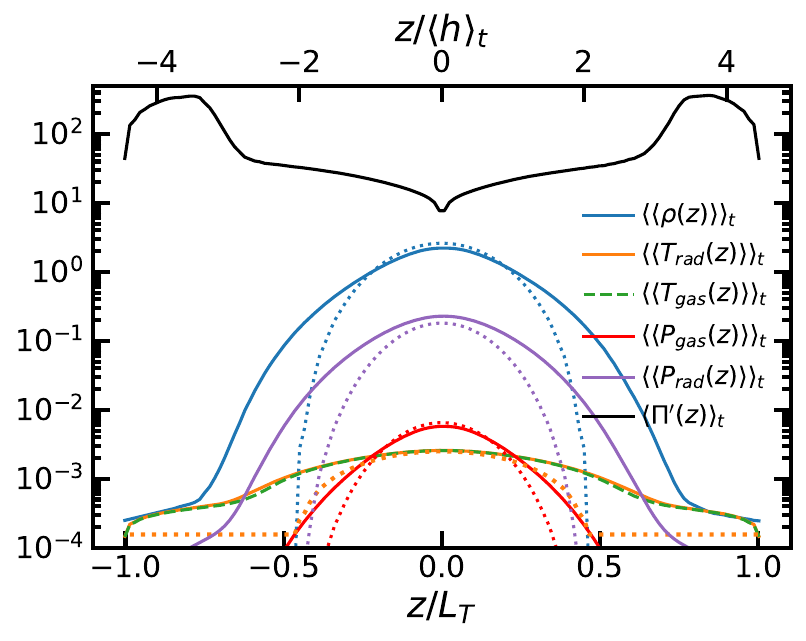}
\caption{The horizontally and temporally 
averaged vertical profiles of density, gas/radiation temperature, and gas pressure in code units for the radiation-dominated run in quasi-steady state. }
\label{fig:vertical_radiation}
\end{figure}

Evolution of globally averaged variables are shown in \autoref{fig:marginalvariable} (analogous to \autoref{fig:fiducialvariable}). 
The run stabilizes after $\sim 100 \Omega^{-1}$, 
converging to $\langle \tau_{\rm cool}\rangle_t \approx 150$, 
$\langle \alpha \rangle_t \approx 0.006$ and $\langle\prg'\rangle_t\approx 38$. 
Other measurements are provided in \autoref{tab:para_turb}.
The averaged vertical profiles in code units are plotted in \autoref{fig:vertical_radiation} in comparison to initial conditions, 
showing relatively extended distributions with a smaller $\langle h \rangle_t/L_T = 0.23$, 
similar to the case with significant radiation \texttt{S2e6O1e-7\_z} shown in \autoref{fig:verticalall}. 
Although for all other cases in \autoref{tab:para_turb} a trend of $\langle h \rangle_t/L_T$ and $\langle Q_T \rangle_t$ increasing with $\langle \prg' \rangle_t$ 
is manifested along the stability boundary, 
a larger $\langle Q_T \rangle_t$ is not seen in run \texttt{S2e6O2e-8\_res}, 
which may be because \texttt{S2e6O2e-8\_res} is not a boundary case.  This  
suggests that for $\prg'\sim 10-30$  there is still room to stabilize at lower $\tau_{\rm cool}$. 

Generally, 
the existence of cases like \texttt{S2e6O2e-8\_res} do suggest that once cooling is weak enough, 
we can end up on a branch of gravito-turbulent states with low turbulence, 
on the other side of the $(\Sigma, \Omega)$ parameter space. 
However, 
from \autoref{fig:boundaryalpha_Prg} we expect such quasi-steady states to have $\alpha\lesssim 0.02$, 
comparable to if not smaller than $\alpha_{\rm MRI}$, 
therefore the significance of such cases is limited 
in a realistic disk environment.

\subsection{Connection to Slow and Rapid Diffusion}
\label{sec:diffusion}

The general behavior of linear gravitational instability in a radiation-pressure-dominated environment has been discussed by \citet{Thompson2008}, who takes into account the destabilizing influence of radiative diffusion.
Particularly relevant here is his analysis for a homogeneous, optically-thick medium in uniform rotation (his Appendix A).
That analysis carries over to the shearing box if restricted to axisymmetric modes ($k_y=0$), provided that the total wavenumber $k=\sqrt{k_x^2+k_z^2}$ is larger than the reciprocal of the vertical density scale height ($H$).
For axisymmetric modes, radial shear influences stability only through its effect on the epicyclic frequency; since the latter is $\kappa=2\Omega$ for uniform rotation but $\kappa=\Omega$ in a Keplerian shearing box, we replace $\Omega$ with $\Omega/2$ in Thompson's formulae.
In particular, Thompson's $Q\equiv\Omega^2/\pi G\rho$ corresponds to $Q/2$ for us.

In the radiation-pressure-dominated regime where $\Pi\gg 1$ but $Q\gtrsim 1$, it is possible to find wavenumbers $k>H^{-1}$ such that $c_{s,gas}^2k^2<4\pi G\rho < c_s^2 k^2$; here $c_{s,gas}^2 = P_{\rm gas}/\rho=c_s^2/\Pi$ is the squared sound speed based just on the thermal pressure.
Under these conditions, \citet{Thompson2008}'s fifth-order dispersion relation (A2) {\it always has at least one unstable root}.
The unstable modes have $k_z\ne0$, which undercuts the stabilizing influence of the epicyclic frequency.
In other words, when $P_{\rm rad}\gg P_{\rm gas}$, 
our shearing boxes are always formally axisymmetrically unstable on scales smaller than the vertical scale height but larger than the isothermal Jeans length $L_{J,\rm gas} = 2\pi/k_{J,\rm gas}$ for  $k_{J,\rm gas}=\sqrt{4\pi G\rho}/c_{\rm s,gas}$.

The small-scale instability depends, however, upon radiative diffusion, 
so that density contrasts can grow without perturbing the radiation pressure.
Thompson distinguishes regimes of slow vs. rapid growth via the dimensionless parameter
\begin{equation}
    \chi \approx \dfrac{ck^2}{3\kappa \rho (4\pi G\rho )^{1/2}}\,,
    \label{eqn:chi_k}
\end{equation}
which compares the rate of diffusion at wavenumber $k$ to the dynamical frequency $\sqrt{4\pi G\rho}$.
We shall use the notations $\chi_{J}$ and $\chi_{J,\rm gas}$ for $\chi$ evaluated at the wavenumbers $k=2\pi/L_J = \sqrt{4\pi G\rho}/c_s$ (i.e., the Jeans wavenumber based on total pressure) and $k=k_{J,\rm gas}$ (i.e. based on thermal pressure), respectively.
Under the above-mentioned conditions where the small-scale mode exists, its growth rate is comparable to the dynamical frequency when $\chi_{J,\rm gas}\gg 1$ 
(i.e. short diffusion time over the gas Jeans length), but slower by a factor $\sim\chi_{J,\rm gas}$ in the opposite limit.

%

\begin{figure}[htbp]
\centering
\includegraphics[width=0.44\textwidth,clip=true]{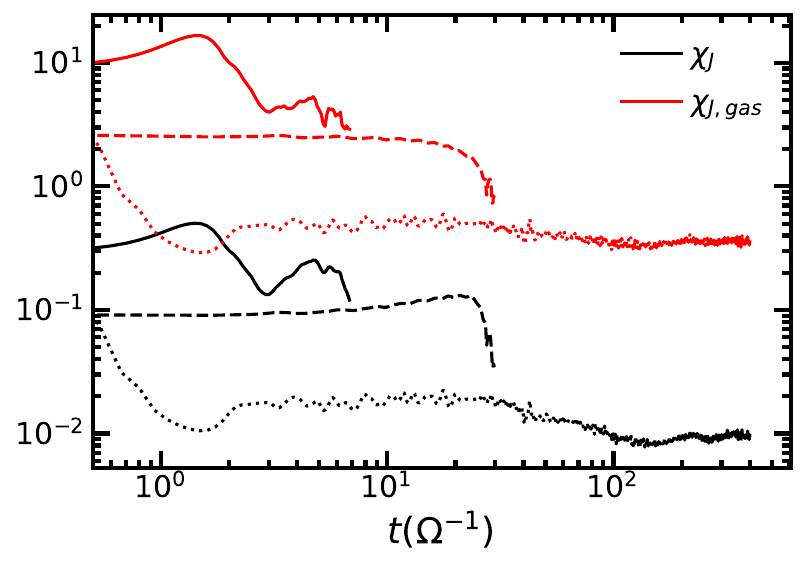}
\caption{The diffusion parameter Equation \eqref{eqn:chi_k} at the Jeans scale based on total pressure ({\it black}) and gas pressure only ({\it red}) for radiation-pressure-dominated runs \texttt{S1e5O2e-9} ({\it solid}), \texttt{S3e5O4e-9} ({\it dashed}), \texttt{S2e6O2e-8\_res} ({\it dotted}).}
\label{fig:chi_summary}
\end{figure}

It can be shown from \autoref{eqn:t_cool_radiation} that $\chi_J\sim\tau_{\rm cool}^{-1}$ when $Q\sim1$.
In \autoref{fig:chi_summary} we plot the time evolution of $\chi_J$ and $\chi_{J,\rm gas}$ from three radiation-pressure-dominated runs.
Run \texttt{S1e5O2e-9} diffuses rapidly even on the scale of the disk thickness ($\chi_J\sim 1$ at its maximum); for this model $\tau_{\rm cool}$ plunges below unity (\autoref{fig:S1e5O18e8variable}), and the disk quickly fragments (\autoref{fig:S1e5O18e8snapshot}, lower row).
Run \texttt{S2e6O2e08\_res} illustrates the opposite extreme, where diffusion on the Jeans length defined by the gas pressure is nearly entering the slow regime ($\chi_{J,\rm gas}\sim 1$), and $\tau_{\rm cool}\gg 1$ as $\chi_{J}\ll 1$ (\autoref{fig:marginalvariable}).
At the end of this run ($t=400\Omega^{-1}$) one sees fine and mildly nonlinear density filaments that plausibly result from the axisymmetric small-scale instability in the regime of slow diffusion where $\chi_{J,\rm gas} < 1$ (\autoref{fig:radiation_snapshot}).
It is possible that this case might eventually fragment if it were continued to longer times.

If non-fragmentation is associated with a slow diffusion even on the gas Jeans scale ($\chi_{J, gas}\lesssim 1$), this requirement roughly translates to $\tau_{\rm cool} \gtrsim \prg$, equivalent to having the cooling time of \textit{gas} thermal energy (instead of gas+radiation energy) longer than the dynamical timescale, in the radiation pressure dominated regime. 

Purely from the numerical simulations, we do not currently have sufficient data to verify whether the critical cooling time in the radiation dominated regime does converge to some power law in $\prg'$ in the limit of large $\prg'$.  In particular, we note that the exemplary simulation \texttt{S2e6O2e-8\_res} is {\it not} a marginally stable case, and lower $\tau_{\rm cool}$ may still be allowed. However, we re-emphasize that as far as physical situations are concerned, we expect the critical cooling time to be at least $\gtrsim 50$ for radiation-dominated  $(\prg' > 1)$ gravito-turbulence (\autoref{fig:boundaryalpha_Prg}), where in an AGN accretion disk context MRI can already provide sufficient heating to suppress gravitational instability.

{It is also possible that the small-scale modes might be entirely suppressed by the introduction of additional physics. In particular, support provided by a magnetic field, unlike that provided by thermal pressure, would not be undercut by radiative diffusion, and the attendant magnetorotational turbulence might disrupt filaments even if they formed.}

{However, non-self-gravitating simulations by \citet[][and references therein]{Turner+2003} have shown that the combination of MRI and radiation pressure can \emph{enhance} density contrasts, provided that the radiation is able to diffuse across an MRI wavelength on timescales $\sim\Omega^{-1}$, 
leaving only the gas pressure to resist compression by the turbulent magnetic field.  
Simulations of galactic disks also show that with moderate-strength magnetic fields (but without radiation), MRI enhances gravitational instability because it undermines stabilization by Coriolis forces \citep{2002ApJ...581.1080K,2003ApJ...599.1157K}. 
Other recent studies in the protoplanetary disk context suggest that MRI induces additional small-scale modes in gravito-turbulent disc regions which could lead to formation of fragments significantly smaller than in GI-only simulations \citep{Deng2020,Deng2021}.
Simulations with self-gravity, radiation, \textit{and} magnetic field will probably be needed to explore the interaction between all these effects that are potentially relevant to AGN accretion disks.}

\section{Discussion}
\label{sec:discussion}
\subsection{Implication for Accretion Disk Structure}
In our gas pressure dominated simulations, we confirm the maximum turbulent stress that quasi-steady gravito-turbulence can support is $\alpha \sim 0.2$, against a cooling timescale of $\tau_{\rm cool} \sim 3-5$. This suggests that in the low $\dot{M}$, $\prg \ll 1$ regime, a quiescent $Q_T\sim 1$ AGN accretion disk heated by gravito-turbulence can extend from the self-gravitating radius $r_{\rm sg}$ of the inner MRI-heated standard disk, towards an outer boundary where the radiative cooling timescale drops to $\tau_{\rm cool}\sim 4\Omega^{-1}$, beyond which fragmentation and star formation dominates.

At high accretion rates, however, the self-gravitating region of an AGN accretion disk is radiation dominated. For a standard disk with turbulent parameter $\alpha_{\rm MRI}$ heated by MRI, the outer boundary $r_{\rm sg}$ where $Q_T$ drops below 1 can be expressed in terms of the accretion rate $\dot{M}$ and $\alpha_{\rm MRI}$.
This $r_{\rm sg}(\dot{M})$ relation should coincide with the mapping between contours of $\dot{M}$ and their \textit{intersection} with the $\tau_{\rm cool} \sim \alpha_{\rm MRI}^{-1} \sim 50$ contour in our $Q_T=1$ plane e.g. \autoref{fig:contours_Mdot}, which shows that  $\prg$ at the intersection $r_{\rm sg}(\dot{M})$ always increases with $\dot{M}$, and $\prg (r_{\rm sg}) \gtrsim 1$ when $\dot{M} \gtrsim M_{\odot}/{\rm yr}$.

More quantitatively, the radiation fraction $\prg$ at $r_{\rm sg}$ satisfies \citep[][modified from their Equation 18]{GoodmanTan2004}
\begin{equation}
\begin{aligned}
    & \left[1+\prg( r_{\rm sg})\right]^{-1/4} \prg( r_{\rm sg})^{-3/4}= 0.416 \left(\dfrac{\alpha_{\rm MRI}}{0.02}\right)^{1 / 3} \\
     &\times\left(\dfrac{\kappa}{0.4 {\rm cm^2/g}}\right)^{-1/2} \left(\dfrac{\mu}{0.6m_p}\right)^{-1} \left(\dfrac{\dot{M}}{2.2 M_\odot {\rm /yr}}\right)^{-1}
\end{aligned}
\end{equation}
which is consistent with the critical accretion rate from our setup. 


As the steady-state $ \prg $ increases in our suite of gravito-turbulent simulations along the fragmentation boundary (see \autoref{tab:para_turb}， \autoref{fig:tcool_data} and \autoref{fig:boundaryalpha_Prg}), due to the increasing destabilizing influence of radiation pressure, $ \tau_{\rm cool} $ continues to increase while $\alpha$ decreases, until for $\prg  \gtrsim 1$, the minimum $\tau_{\rm cool}$ reaches $\sim 50$. This suggests that Eddington (for $M_{\rm SMBH}\lesssim 10^8M_\odot$) or near-Eddington 
(for $M_{\rm SMBH}\sim 10^9M_\odot$)
radiation dominated AGN accretion disks simply cannot have a quasi-steady gravito-turbulent region beyond $r_{\rm sg}$, 
since gravito-turbulence could not provide significantly larger heating compared to MRI.  This suggests that the disk directly switches to a star-forming region without a transition zone that has quasi-steady gravitationally driven turbulence and accretion. 

In principle, the energy requirements to maintain thermal equilibrium in the star-forming region can be maintained by star formation feedback, similar to the overall situation in star-forming galactic disks \citep{2010ApJ...721..975O,2011ApJ...731...41O, 2022ApJ...936..137O}, and this need not require $Q_T\sim 1$ since turbulent dissipation and rapid cooling allow collapse to occur at scales $\lesssim H$.
To continue feeding gas to the inner disk,  $\alpha$ is maintained to be some maximum value that is $\sim 0.3$ for $\prg \ll 1$, or smaller than $\lesssim 0.02$ for radiation pressure dominated disks, 
although decoupled from the energy equation. 
Alternatively, 
the accretion radial velocity may be parameterized by $\dot{M}/2\pi r \Sigma =v_r=mh \Omega r$ instead of $\sim \alpha h^2 \Omega r$, 
if accretion is mainly driven by global instabilities \citep{TQM} or large-scale magnetic torques (Sanghyuk Moon et al 2023, ApJ submitted).


\subsection{Implication for Stellar Evolution in AGN Disks}

In most fragmenting cases, 
the fragments' \textit{initial} masses are of order the Jeans mass, 
\begin{equation}
    M_J = \Sigma L_J^2 \approx  \Sigma \left(\dfrac{\pi^2 G \Sigma}{\Omega^2}\right)^2 \approx 20 M_\odot \Sigma^{3}_{-5} \Omega^{-4}_{-8}, 
\end{equation}
as predicted by classical theories and as found in numerical simulations of gravitationally-unstable disks \citep[e.g.][]{2002ApJ...581.1080K,2003ApJ...599.1157K}
For $t_{\rm cool} \ll \Omega^{-1}$, in the radiation pressure dominated case (see \S \ref{sec:rad_frag_results}), 
we find that radiative diffusion can be very fast and the fragments instead  have masses of
\begin{equation}
    M_{J, \rm gas } = \Sigma \dfrac{L_J^2}{\prg} \approx  \Sigma \left(\dfrac{\pi^2 G \Sigma}{\Omega^2 }\right)^2\dfrac1{\prg} \approx 2000 M_\odot \Sigma^{1.5}_{-5} \Omega^{-2}_{-9}.
\end{equation}
Since in the radiation pressure dominated regime $\prg \propto \Sigma^{1.5} \Omega^{-2}$  (\autoref{eqn:T_prop}), 
the initial masses can still be very large for super Eddington AGN disks with high radiation pressure fraction. 
Nevertheless, the initial masses of these fragments may not be directly relevant to their final masses, 
since through either collisions or gas accretion they may quickly grow towards a mass limit constrained by Hill radius isolation  
\citep{GoodmanTan2004} or Eddington limit \citep{Cantiello2021,Jermyn2022}.

Generally, 
our simulation outcomes strongly favor star formation in radiation pressure dominated AGN disks. 
The rapid accretion and pollution of disk gas by the massive stars that are formed here, 
and possibly their eventual supernovae, may contribute to super-solar metallicity abundances in AGNs \citep{Hamann1999,Hamann2002}.
The long-term evolution of massive stars may leave behind embedded stellar mass black holes (EBHs) that could provide extra heating through accretion feedbacks \citep{GilbaumStone2022}, and also relevant to production of gravitational waves (GW) that may contribute 
to LIGO-Virgo events \citep{McKernan2012, McKernan2014,Stone2017,Tagawa2020a,Samsing2022,Li2022}. 
In particular, 
BBH mergers in an 
AGN disk may produce electromagnetic counterpart that could 
differentiate them from other merger channels \citep{graham2020}. 

\subsection{Future Prospects}

Given adequate computational resources, it will be possible to perform a more extensive parameter survey over $(\Sigma, \Omega)$ space, 
in order to map out a more complete $ \prg  - \tau_{\rm cool}$ scaling that extends far into the radiation dominated regime on \autoref{fig:tcool_data}. 
Nevertheless, with MRI present as an auxiliary heating source, our finding of a $\tau_{\rm cool} \gtrsim 50$ constraint up to $\prg \sim 1$ is sufficient evidence to declare our major conclusion: super Eddington accretion disks will always fragment beyond $r_{\rm sg}$, 
and $\tau_{\rm cool}\gtrsim 50, \alpha < \alpha_{\rm MRI}$ gravito-turbulent states for even larger $\prg$ are possible (e.g. \S \ref{sec:radiation_turbulence}) but not physically significant except in situations where MRI is suppressed.

A more meaningful direction for further study may be to extend the currently explored parameter space to a more realistic setup. 
In this paper, we have assumed constant frequency-averaged opacity $\kappa$ and a gray model $\kappa_R = \kappa_P$ for convenience in estimating initial cooling times and comparing with analytical disk profiles. 
It would be straightforward to instead implement realistic opacities as functions of density and temperature in subsequent studies. 

A caveat of our study is that we only focus on sketching the fragmentation boundary and do not follow the evolution of the gravitationally bound structures that form. 
Since we expect star formation to generically develop in the outer regions of AGN disks 
(shortly beyond $r_{\rm sg}$ for near-Eddington, radiation dominated cases), 
it will be very interesting to follow this in more detail.  In general, sink particles are needed to avoid numerical singularities from gravitational collapse, and these can be coupled to treatments of radiation feedback, stellar winds, and supernovae given assumptions of the stellar population that is produced  \citep[e.g.][]{2013ApJS..204....8G,2015ApJ...809..187S,2017ApJ...846..133K,2017ApJ...851...93K,2021ApJ...914...90L}.  
With the addition of adaptive mesh refinement to follow small-scale fragmentation, it will be possible to follow the long-term evolution of the initial Jeans-scale clumps that are the dominant type of outcome in our $\prg \gtrsim 1$ models. 
Such simulations, either local or global, 
can inform us both regarding the mass spectrum of stars formed, 
and whether energy equilibrium with star formation heating is indeed maintained, either in a state of 
marginal large-scale gravitational instability
\citep{TQM},
or a state of vertical thermal and dynamical equilibrium more similar to nuclear rings fed by bars at larger scale \citep{2021ApJ...914....9M}.

\acknowledgements

YXC thanks Wenrui Xu, Chang-Goo Kim, Alwin Mao, Douglas Lin, Eliot Quataert, Xue-Ning Bai, Jane Dai, Kaitlin Kratter, Minghao Guo for helpful discussions. We also acknowledge computational resources provided by the high-performance computer center at Princeton University, which is jointly supported by the Princeton Institute for Computational Science and Engineering (PICSciE) and the Princeton University Office of Information Technology. The work of ECO is supported by grant 510940 from the Simons
Foundation. The Center for Computational Astrophysics at the Flatiron Institute is supported by the Simons Foundation.

\appendix
\section{Initial Equilibrium Profile}
\label{appendix}
To set up a fiducial initial profile, we assume  $\Pi := P_{\rm rad}/P_{\rm gas}$ is constant in the bulk of the disk, which suggests a $P(z)=K\rho^{4/3}(z)$ polytropic profile that is analytically solvable given fixed parameters $\Sigma, \Omega$, and an additional $Q$ (or equivalently the midplane density $\rho_0$, see eq.~\eqref{eq:altQ}) that controls the onset of GI \footnote{the polytropic assumption is used only for this initialization; the subsequent evolution obeys the full energy equation}. Although $Q$ is not conserved throughout the simulation, we expect $Q \sim Q_T \sim 1$ if our initial conditions indeed evolve into steady states, consistent with the arguments in \S \ref{sec:formulation}. 

We can express the constant $K$ in terms of $\rho_0$ and $\Sigma$ \citep[][using Equations A2, A5]{Jiang11}:

\begin{equation}
    \mathcal{H}:=\dfrac{\Sigma}{4\sqrt{2} \rho_0 I_3(Q)}=\sqrt{\dfrac{K}{\rho_0^{2/3}\pi G}}
    \label{eqn:scaleheight}
\end{equation}

Where $\mathcal{H}$ is a normalization vertical length and a proxy for the photosphere height. Here we define $I_k(Q)$ as

\begin{equation}
\begin{aligned}
I_{k}(Q) \equiv \int_{0}^{1} \frac{\left(1-w^{2}\right)^{k} d w}{\sqrt{2 Q+1+\sum\limits_{i=1,2,3}\left(1-w^{2}\right)^{i}}}.
\end{aligned}
\end{equation}

Approximations accurate to \% 1 are \citep[][Equation 8]{Jiang11}:

\begin{equation}\label{eq:structure_approximations}
    I_{3}(Q) \approx \frac{0.323}{\sqrt{Q+1.72}}, \quad I_{4}(Q) \approx \frac{0.287}{\sqrt{Q+1.72}},
\end{equation}

Given $\mathcal{H}$ determined by $Q, \rho_0, \Sigma$, the vertical distribution of density $\rho(z)$ is initialized according to

\begin{equation}
\begin{aligned}
    \dfrac{z}{\mathcal{H}} =&
     \int_{0}^{\sqrt{1-\theta}}\frac{2 \sqrt{2}d w}{\sqrt{2 Q+1+\sum\limits_{i=1,2,3}\left(1-w^{2}\right)^{i}}},\\ \theta =& [\rho(z)/\rho_0]^{1/3}
\end{aligned}
\end{equation}

It can be shown that $\rho(z)$ converges to the following analytic relation when $Q$ approaches infinity (non-self-gravitating case):

\begin{equation}
    [\rho(z)/\rho_0]^{1/3}=(1-\dfrac{z^2\Omega^2}{8K\rho_0^{1/3}}).
    \end{equation}

The specific vertical profile with constant $K=P/\rho^{4/3}$ gives coefficients for the midplane EoS $P_0(\rho_0, T_0)$ (\autoref{eq:quartic}) as

\begin{equation}
    \dfrac{\rho_0}{\rho_*} = f_\rho = \dfrac1{2Q}, \dfrac{P_0}{P_*} = f_p = \dfrac1{32[I_3(Q)]^2}
    \label{eq:rho0_P0}
\end{equation}

For vertically integrated pressure $P_{\rm 2D}$ and internal energy density $U_{\rm 2D}$, Equations 5 \& 9 of \citet{Jiang11} translate to

\begin{equation}
    \dfrac{P_{\rm 2D}}{P_* l_*}=\frac{Q I_4(Q)}{16\left[I_3(Q)\right]^3}, U_{\rm 2D} = \left[1-\frac{1}{2(\Pi+1)}\right] P_{\rm 2D},
    \label{eqn:P2D}
\end{equation}

which gives us another order-unity coefficient

\begin{equation}
    \dfrac{U_{\rm 2D}}{P_* l_*} = f_U = \left[1-\frac{1}{2(\Pi+1)}\right] \frac{Q I_4(Q)}{16\left[I_3(Q)\right]^3}
    \label{eqn:U2D}
\end{equation}

Apart from hydrodynamical variables, we also attempt to set up initial vertical radiation flux self-consistently. Since $\Pi$ is independent of $z$, from the expression for optically thick radiation pressure gradient

\begin{equation}
    \dfrac{dP_{\rm rad}}{dz}= \Pi \dfrac{dP_{\rm gas}}{dz} = - \kappa \rho F_z/c
\end{equation}


Where 
\begin{equation}
    \dfrac{dP_{\rm rad}}{dz}+ \dfrac{dP_{\rm gas}}{dz} = -\Omega^{2} z-4 \pi G \int_{0}^{z} \rho\left(z^{\prime}\right) d z^{\prime}
\end{equation}

We can calculate the initial $F_z$ distribution with the opacity $\kappa = \kappa_s+\kappa_R$. As one go to higher $|z|$, $\rho$, $T$ as well as $E_{rad}$ start to abruptly drop at $\sim \pm \mathcal{H}$, which is also where the cumulative optical depth falls to unity, so we fix the value of both $E_{rad}$ and $F_z$ outside this ``photosphere" around $\sim \pm \mathcal{H}$ where $|F_{z,\rm max}| /c = E_{rad} = a T_{\rm eff}^4$ are all constants. This means regions outside photosphere does not have fixed $\Pi$ anymore and is subject to optically thin cooling. Nevertheless, energy distribution outside the photosphere has little relevance to the turbulence process since gas is tenuous there. We find that as long as $|F_z| /c \lesssim E_{rad}$ beyond $\mathcal{H}$, the radiation profile outside the photosphere relaxes to a constant $F_{z,\rm max}:=a T_{\rm eff}^4$ after a small timestep, which is close to the analytical prediction given in \autoref{eq:Fzmax}. With the numerical solution for $F_{z,\rm max}$, we can plot out scalings for the expected cooling timescales and local accretion rates (\autoref{fig:contours_tcool} and \autoref{fig:contours_Mdot}).

\bibliographystyle{aasjournal}
\bibliography{main}

\end{CJK*}
\end{document}